\documentclass[%
reprint,
superscriptaddress,
 amsmath,amssymb,
 aps,
]{revtex4-2}

\usepackage{graphicx}% Include figure files
\usepackage{dcolumn}% Align table columns on decimal point
\usepackage{bm}% bold math

\usepackage{ulem}
\usepackage{amsmath}

\usepackage{xcolor}

\begin{document}

\preprint{APS/123-QED}

\title{Dissipation reduction and information-to-measurement conversion in DNA pulling experiments with feedback protocols}

\author{M. Rico-Pasto}
 \thanks{Equally contributed authors}
 \affiliation{%
 Condensed Matter Physics Department, University of Barcelona, C/Marti i Franques 1, 08028 Barcelona, Spain}%
 
\author{R.K. Schmitt}
 \thanks{Equally contributed authors}
 \affiliation{%
 Solid State Physics and NanoLund, Lund University, PO Box 118, SE-221 00 Lund, Sweden}%
  
\author{M. Ribezzi-Crivellari}
 \affiliation{%
 École Supérieure de Physique et de Chimie Industrielles | ESPCI · Laboratoire de Biochimie (LBC), 75005 Paris, France}%

\author{J.M.R. Parrondo}
 \affiliation{%
 Departamento de Estructura de la Materia, Física Térmica y Electrónica and GSIC, Universidad Complutense de Madrid, 28040 Madrid, Spain}%

\author{H. Linke}
 \affiliation{%
 Solid State Physics and NanoLund, Lund University, PO Box 118, SE-221 00 Lund, Sweden}%

\author{J. Johansson}
 \affiliation{%
 Solid State Physics and NanoLund, Lund University, PO Box 118, SE-221 00 Lund, Sweden}%

\author{F. Ritort}
\email[Corresponding author: ]{ritort@ub.edu}
 \affiliation{%
 Condensed Matter Physics Department, University of Barcelona, C/Marti i Franques 1, 08028 Barcelona, Spain}%
 \email{fritort@gmail.com}

\date{\today}

\begin{abstract}
Information-to-energy conversion with feedback measurement stands as one of the most intriguing aspects of the thermodynamics of information in the nanoscale. To date, experiments have focused on feedback protocols for work extraction. Here we address the novel case of dissipation reduction in non-equilibrium systems with feedback. We perform pulling experiments on DNA hairpins with optical tweezers, with a general feedback protocol based on multiple measurements that includes either discrete-time or continuous-time feedback. While feedback can reduce dissipation, it remains unanswered whether it also improves free energy determination (information-to-measurement conversion). We define thermodynamic information $\Upsilon$ as the natural logarithm of the feedback efficacy, a quantitative measure of the efficiency of information-to-energy and information-to-measurement conversion in feedback protocols. We find that discrete and continuous-time feedback reduces dissipation by roughly $k_BT \Upsilon$ without improvement in free energy determination. 
Remarkably, a feedback strategy (defined as a correlated sequence of feedback protocols) further reduces dissipation, enhancing information-to-measurement efficiency. Our study underlines the role of temporal correlations to develop feedback strategies for efficient information-to-energy conversion in small systems.
\end{abstract}

%Use showkeys class option if keyword display desired

\maketitle

\section{ Introduction}\label{sec.Intro}

Maxwell’s demon (MD) thought experiment \cite{Leff1990} has led to the insight that information enhances the capacity to extract energy from a system. In 1961 Landauer demonstrated that any irreversible logical operation (such as bit erasure) dissipates at least $k_BT\log{2}$ per stored bit of information \cite{Land1961}. Bennett \cite{Ben1973} applied this result to the Szilard’s engine \cite{Leff1990} – a single particle version of the MD that extracts energy
from a thermal bath in a cycle– and showed that bit erasure is an entropy producing step necessary to restore the initial \textit{blank} state of the memory of the demon, in agreement with the second law \cite{Ben1982,Maru2009,Lutz2015}. These developments have boosted a new field of research, namely the thermodynamics of small systems under feedback control \cite{Tou2000,Cao2009,Sag2014,Par2015}. This has led to experimental realizations of the Maxwell demon in colloidal systems \cite{Toy2010,Pan2018,Pan2018_2,Adm2018},  electronic \cite{Kos2014,Kos2014_2} and optical devices \cite{Vid2016,Chi2017,Kum2018}, single molecules \cite{Rib2019}, quantum systems \cite{Cot2017,Mas2018,Nag2018},  and tests of the Landauer limit \cite{Ber2012,Ber2013,Jun2014,Hong2016,Pet2016}. The extension of stochastic thermodynamics \cite{Sei2012,Ale2015,Cili2017} to include information and feedback has led to novel fluctuation theorems (FTs) for work and information \cite{Kim2007,Sag2010,Ponm2010,Abreu2012}, in repeated-time feedback protocols \cite{Horo2010,Sag2012,Ash2014,Lah2016} and analytically solvable models \cite{Man2012,Stra2013,San2019,Ann2020}. Generalized Jarzynski equalities have been derived for isothermal feedback processes, where an external agent performs a single measurement on a system and applies a protocol $\omega_m$ that depends on the measurement outcome $m=1,2,\ldots,M$. A main equality useful for measurements with feedback reads \cite{Sag2010},
\begin{equation}\label{eq.Jarz}
    \langle \exp \left[ -(W-\Delta G)/k_BT\right] \rangle = \sum_{m=1}^{M}P_{\leftarrow}(m|\omega_m) \equiv \gamma
\end{equation}
with $k_B$ the Boltzmann constant and $T$ the temperature. Here, $W$ is the work performed on the thermodynamic system, $\Delta G$ is the free energy difference and $P_{\leftarrow}(m|\omega_m)$ is the probability to measure $m$ along the time reversal ($\leftarrow$) of the original protocol $\omega_m$. Equation \eqref{eq.Jarz} defines the efficacy parameter $\gamma(\le M)$ which quantifies the reversibility of the feedback process \cite{Horo2011}, reaching its maximum value ($M$) for reversible feedback process where $P_\leftarrow(m|\omega_m)=1$ for all $m$. Without feedback $\omega_m\equiv\omega$ and $\gamma=1$, with Eq.\eqref{eq.Jarz} the Jarzynski equality. It is convenient to define the logarithm of the efficacy: $\Upsilon=\log{\gamma}$, which is a bound of the work that can be extracted in an isothermal feedback process. $\Upsilon$ might be called thermodynamic information, information utilization or negentropy. For discrete-time single measurements $\Upsilon$ is bounded from above (Eq.\eqref{eq.Jarz} with $P_{\leftarrow}(m|\omega_m)\le 1$), $-\infty<\Upsilon\le\log{M}$ ($M=2$ being the one-bit Landauer limit). Jensen’s inequality applied to Eq.\eqref{eq.Jarz} yields, 
\begin{equation}
 \langle W_d \rangle \equiv \langle W \rangle -\Delta G \geq -k_BT \Upsilon\Rightarrow \langle W_d \rangle+k_BT \Upsilon\geq 0
 \label{eqWdis}   
\end{equation}
where $\langle W_d \rangle$ is the dissipated work. Without feedback $\Upsilon = 0$ and $\langle W_d \rangle_0 \geq 0$ where the subscript $0$ denotes the non-feedback case. In experimental realizations of the MD, feedback measurement is operated in equilibrium systems and $\Upsilon=\log{M}$ is the maximum extractable (i.e., negative) work for equally likely outcomes, $\langle W \rangle = -k_BT\Upsilon = -k_BT\log M$. 

Here we address a novel situation where feedback is operated to reduce dissipation in small non-equilibrium systems. Most systems in nature and daily life applications are out-of-equilibrium and dissipative, requiring feedback to reduce dissipation. Examples range from heat engines that minimize heat dissipation to control energy production and avoid extreme events (e.g., accidents in power plants) to living organisms.  Most regulatory processes in the cell focus on housekeeping tasks: the entropy production must be kept within bounds and reduced upon unexpected rises due to exogenous (external) factors. The understanding of dissipation reduction is vital in the nanoscale where dissipative molecular processes are remarkably efficient (for instance, the typical $>90\%$ efficiency of the ATPase machinery). In all these non-equilibrium systems, dissipation reduction by using feedback is critical. The main goal of this paper is to derive and test fluctuation theorems describing dissipation reduction in such kinds of systems.

For a non-equilibrium process Eq.\eqref{eqWdis} shows that dissipation is bounded by $-k_BT\Upsilon$. For $\Upsilon>0$ (information-to-work conversion) dissipation is reduced by at most $-k_BT\Upsilon$. Conversely, one could apply feedback protocols where $\Upsilon<0$ (information-to-heat conversion) and dissipation increased by at least $-k_BT\Upsilon$. The latter case is non-productive feedback for dissipation reduction. It has similarities with feedback in control theory, where protocols regulate experimental variables, e.g., by keeping them constant \cite{bechhoefer2005,dieterich2016}. These types of protocols counteract deviations from a system’s specific pre-set conditions rather than rectifying thermal fluctuations, leading to increased dissipation. For the relevant case $\Upsilon>0$ the dissipated work is reduced with respect to the non-feedback case, $\langle W_d \rangle \leq \langle W_d \rangle_0$. We define the information-to-energy or feedback cycle efficiency \cite{Ver2014,Sch2015,Rib2019_2}, $\eta_I$, as the reduction in dissipation, $\Delta \langle W_d\rangle=\langle W_d \rangle_0 - \langle W_d\rangle$, relative to the second law bound, $\langle W_d \rangle_0 + k_BT \Upsilon$: 
\begin{equation}\label{eq.etaI}
\eta_I = \frac{\Delta \langle W_d\rangle}{\langle W_d \rangle_0 + k_BT\Upsilon} < 1.
\end{equation}
This is our first key result for information-to-work conversion in non-equilibrium conditions. For cyclic and reversible MD devices $\langle W_d \rangle_0 = 0$ so $\eta_I = -\langle W_d \rangle / k_BT \Upsilon$ is the standard MD efficiency.

Related to dissipation reduction is free energy determination, a relevant question for molecular thermodynamics where important applications have emerged in molecular folding and ligand binding \cite{Juni2009,Ale2012,Camu2017}. It is an open question whether, by reducing dissipation, feedback can improve free energy determination, what we call information-to-measurement conversion. Free energy determination can be improved if $\langle W_d \rangle+k_BT \Upsilon<\langle W_d \rangle_0$, which we denote as weakening of the second law. Free energy determination improvement is related to the Jarzynski relation Eq.\eqref{eq.Jarz} and its bias, $B_N$, for $N$ work ($W$) measurements. Inserting $\Upsilon=\log{\gamma}$ in Eq.(\ref{eq.Jarz}) we define, 
\begin{eqnarray}
    & &B_N= \langle \Delta G_N\rangle-\Delta G\nonumber\\
    & &\Delta G_N=-k_BT \log{ \left( \frac{1}{N}\sum_{i=1}^{N} e^{-\frac{W_i}{k_BT}} \right)}+k_BT\Upsilon 
\label{eqBias}
\end{eqnarray}
with $\Delta G_N$ the Jarzynski $\Delta G$-estimator for $N$ experiments and $\langle..\rangle$ the average over many realizations of the $N$ experiments. The exponential average of minus the work in Eq.\eqref{eqBias} is biased for finite $N$ (whereas the bounded and finite sum defining $\Upsilon(=\log\gamma)$ in Eq.\eqref{eq.Jarz} is not).  $B_N$ is positive and monotonically decreasing with $N$ \cite{Pala2011}, vanishing in the limit $N\to\infty$. Therefore improved free energy determination requires that $B_N$ for $N=1$ decreases with feedback relative to the non-feedback case. From Eq.(\ref{eqBias}) we have, 
\begin{eqnarray}
      B_1 &=& \langle \Delta G_1\rangle-\Delta G =\langle W\rangle-\Delta G+k_BT\Upsilon= \\
      & & = \langle W_d\rangle+k_BT\Upsilon\ge 0 \nonumber
\label{eqBiasN1}  
\end{eqnarray}
which is equivalent to Eq.\eqref{eqWdis}. Therefore, weakening of the second law implies reducing $B_1$, leading to improved free energy determination for finite $N$.

To quantify information-to-measurement conversion we define the cycle efficiency $\eta_M$, as the relative difference between the second law inequality bounds with feedback, $\langle W_d\rangle + k_BT\Upsilon \geq 0$, and without feedback, $\langle W_d\rangle_0 \geq 0$: 
\begin{equation}\label{eq.etaM}
\eta_M =1-\frac{\langle W_d\rangle + k_BT\Upsilon}{\langle W_d\rangle_0}= \eta_I + (\eta_I - 1)\frac{k_BT\Upsilon}{\langle W_d \rangle_0}.
\end{equation}
This is our second main result, which leads to a new inequality, $\eta_M  \leq \eta_I$. Note that $\eta_M = 1$ if and only if $\eta_I = 1$, in which case dissipation reduction is maximal, $\langle W_d \rangle = -k_BT\Upsilon$, and $\Delta G = \langle W \rangle + k_BT\Upsilon$ can be determined with certainty. Improved free energy determination requires $\eta_M > 0$, whereas for $\eta_M \leq 0$ no gain in free energy determination is obtained with feedback: $\langle W_d \rangle$ decreases with respect to $\langle W_d \rangle_0$ by exactly or less than $k_BT\Upsilon$. In general, optimal free energy determination is obtained by maximizing $\eta_M$.

%_________________________________________________
%%%%%%%%%%%%%%%%%%%%%%%%%%%%%%%%%%%%%%%%%%%%%%%%%%
% _____________________________________ FIGURE 1 
\begin{figure*}
\centering\includegraphics{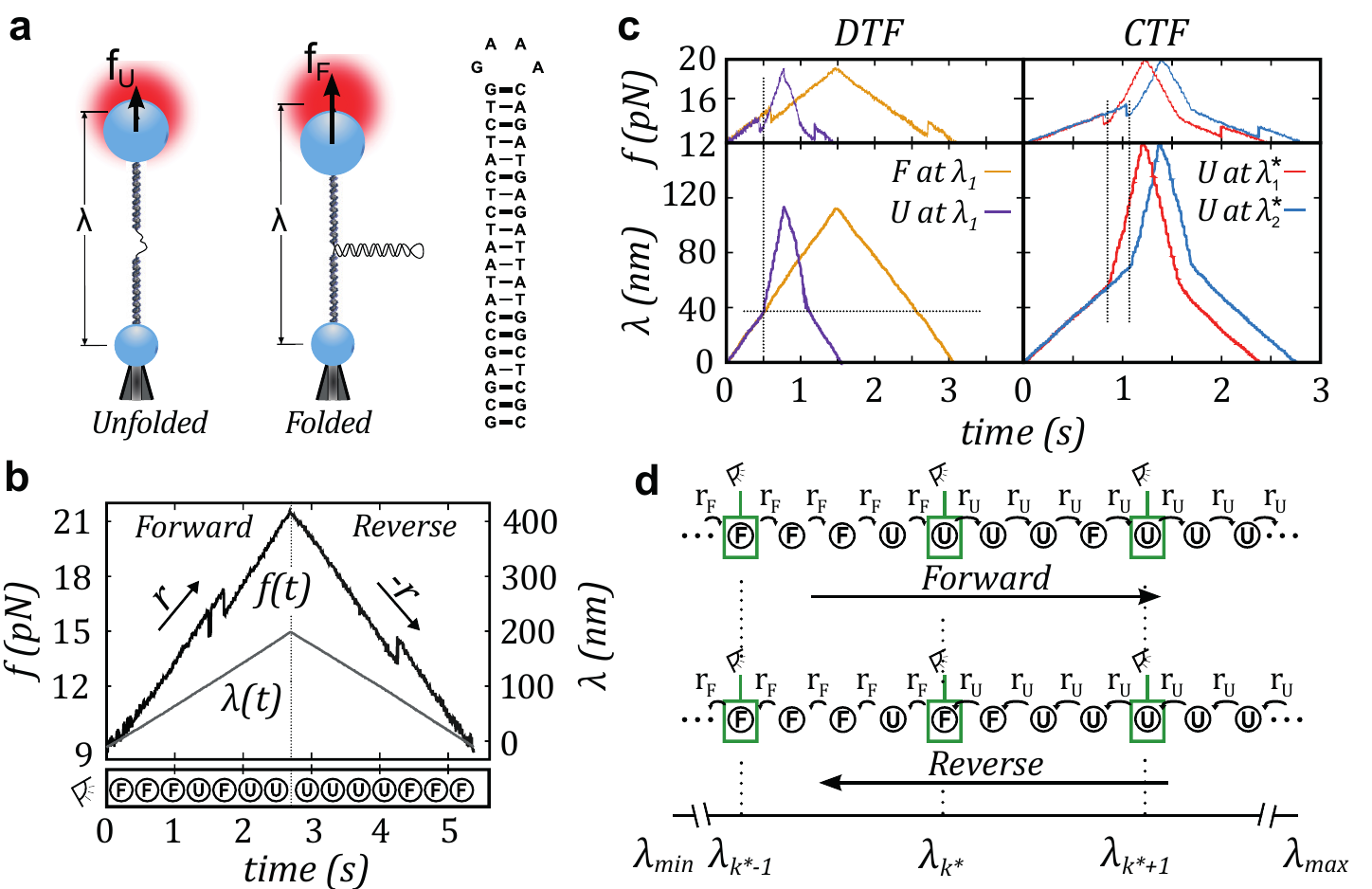}
\caption{\label{fig1} {\bf DNA pulling experiments with feedback measurement} (a) Schematics of the single-molecule experimental setup. The optical trap measures the force $f$ applied on the molecule. The control parameter $\lambda$ denotes the distance between the optical trap and the pipette. The DNA hairpin sequence is shown on the right. (b) Experimental unfolding/refolding FDCs without feedback, where we plot both $f$ and $\lambda$ as a function of time. $\lambda$ (color gray, scale on the right side) is first increased (unfolding/forward process) at a constant loading rate $r=4$pN/s. After reaching a predefined upper value, $\lambda$ is decreased at the unloading rate $r$ (folding/reverse process). Along the time axis we show the sequence of states (F, folded; U, unfolded) observed at specific times along the trajectory (note the transient refolding event along $\rightarrow$). (c) Left: Pulling protocol with DTF. The initial pulling rate $r_F$ is switched to $r_U$ (purple) if the molecule is found to be unfolded (U) at $\lambda_1$; and is unchanged otherwise (orange). Right: Pulling protocol with CTF. The pulling rate $r_F$ along $\rightarrow$ is switched to $r_U$ as soon as an unfolding event is detected. In both cases the protocol in the $\leftarrow$ process is the time-reversal of the one used in the $\rightarrow$ process. (d) Schematics of the general $1^{st}$time-feedback ($1^{st}$TF) protocol in a pulling experiment. The state of the system (F or U) is repeatedly observed during $\rightarrow$ at specified values of the control parameter $\lambda_k$ (green boxes). At $\lambda_{k^\ast}$ the system is observed to be in U for the first time, and the pulling rate switched from $r_F$ to $r_U$. In the illustration, the first time the molecule is found to be in U along the pre-determined set of $\lambda_k$ occurs at $\lambda_{k^\ast}$: the molecule was in F at all observed trap positions before $\lambda_{k^\ast}$ (with $\lambda_{k^\ast-1}$ the previous trap position where F was observed). Note that the molecule can execute multiple transitions at intermediate values of $\lambda$ (between $\lambda_{k^\ast-1}$ and $\lambda_{k^\ast}$)  where no observations are made. A feedback response is only triggered when state U is observed for the first time at the specific positions defined by the pre-determined set $\lbrace \lambda_k\rbrace$.}
\end{figure*}

Here we address information-to-energy and information-to-measurement conversion by combining theory and experiment. We introduce a new feedback-FT for multiple repeated measurements that is applicable to DNA unzipping experiments with optical tweezers (Fig.\ref{fig1}a and Methods \ref{sec:OT_dev}). From the pulling experiments we measure the work distributions and  $\Upsilon$ to extract the efficiencies $\eta_I,\eta_M$. We investigate whether reduced dissipation with feedback ($\Upsilon>0$) improves free energy determination. In a pulling experiment without feedback the optical trap is repeatedly ramped up (forward,$\rightarrow$) and down (reverse,$\leftarrow$) at a constant speed between trap positions $\lambda_{min}$ and $\lambda_{max}$ where the molecule is folded (F) and unfolded (U), respectively, while the force $f$ exerted by the trap is measured, producing a force-distance (FDC) curve (Fig.\ref{fig1}b and Methods \ref{sec:pullexp}). We apply two different feedback protocols, namely discrete-time feedback (DTF) (Fig.\ref{fig1}c, left) and continuous-time feedback (CTF) (Fig.\ref{fig1}c, right). The CTF protocol is the non-equilibrium generalization of a recently introduced continuous Maxwell demon \cite{Rib2019}, DTF and CTF being particular cases of the new feedback protocol. We will show that, while DTF and CTF protocols mildly reduce dissipation, they do not improve free energy determination ($\eta_M\lesssim 0$). Remarkably, a feedback strategy combining DTF and CTF protocols markedly increases $\eta_I$ and $\eta_M$. This sets feedback strategies (defined as a sequence of multiple-correlated feedback protocols) as the route to enhance the non-equilibrium information-to-energy and -to-measurement efficiencies.

\section{\label{seq.Methods}Materials and methods}
\subsection{Instrument design and single molecule construct}
\label{sec:OT_dev}
The instrument used in this study is a miniaturized optical tweezers setup \cite{Hug2010} with counter-propagating lasers focused into the same point to
create a single optical trap. The control parameter for the device is the position of the optical trap with respect a fixed point, in our case the bead immobilized
on the tip of a micropipette (Fig.\ref{fig1}a). Force is directly measured from the change in light momentum of the deflected beam using position sensitive detectors. The molecular construct is manipulated by using two polystyrene beads, one coated with anti-digoxigenin, the other with streptavidin, that are connected to opposite ends of the molecular construct through antigen-antibody and biotin-streptavidin bonds, respectively.

For the experiments we have used a molecular construct made of a short DNA hairpin linked to molecular handles on both flanking sides. The DNA hairpin has a stem consisting of 20 base pairs (bp) and a tetraloop (underlined), (5’-GCGAGCCATAATCTCATCTG \underline{GAAA} CAGATGAGATTATGGCTCGC-3’) that unfolds and refolds cooperatively in a two-state manner. It is flanked by two identical 29 bp double-stranded DNA (dsDNA) handles \cite{Forns2011}. The molecular handle is tagged with a single biotin at the 3’-end while the other end is tagged with multiple digoxigenins. For pulling experiments the molecular construct (DNA hairpin plus handles) is tethered between two beads, one is captured in the optical trap and the other is immobilized on the tip of a glass micropipette.
\subsection{Pulling experiments and work measurements}
\label{sec:pullexp}
In pulling experiments without feedback a mechanical force is applied to the ends of the molecular construct (Fig.\ref{fig1}a). At low forces, typically below 10pN, the hairpin remains in its native double-stranded configuration (folded state, F), while at higher forces it unfolds to its single-stranded DNA configuration (unfolded state, U). Every pulling cycle consists of two processes (Fig.\ref{fig1}b): the forward (unfolding) process ($\rightarrow$) where the molecule is initially in F at $\lambda_{min}$ and the force $f$ increases at a constant loading rate $r$ until $\lambda_{max}$ is reached. During this process the molecule unfolds, entering state U, and the force drops by ${\Delta f} \sim 1$pN; In the reverse (folding) process ($\leftarrow$) the molecule is initially in U at $\lambda_{max}$ and the force is decreased at the unloading rate $-r$ until reaching $\lambda_{min}$. When the molecule folds, the force rises by the same amount ${\Delta f} \sim 1$pN. For a given trajectory one can extract the work $W$ exerted by the optical trap on the molecular construct. $W$ is defined as the area below the force-distance curve, $W_{\rightarrow (\leftarrow)} = +(-) \int_{\lambda_{min}}^{\lambda_{max}} fd\lambda$ with positive (negative) $W$ values for the $\rightarrow$ ($\leftarrow$) process. The pulling experiment defines a thermodynamic transformation on the molecular system (DNA hairpin, handles and bead) between trap positions $\lambda_{min}$ and $\lambda_{max}$ with free energy difference ${\Delta}G_{FU} = G_U\left(\lambda_{max}\right )- G_F\left(\lambda_{min}\right)$. The second law states that $\langle W\rangle_{\rightarrow(\leftarrow)}\geq+(-)\Delta G_{FU}$, so the average dissipated work fulfils $\langle W_d \rangle_{\rightarrow (\leftarrow)} = \langle W\rangle_{\rightarrow(\leftarrow)} -(+) \Delta G_{FU} \geq 0$, vanishing for a quasi-static process (i.e., infinitely slow pulling or $r \rightarrow 0$). If the loading rate is not too high, the molecule can unfold and refold more than once, the average dissipated work being typically small $\langle W_d\rangle\sim k_BT$. For higher loading rates, the process becomes irreversible and the thermally activated unfolding tends to occur at higher forces while the refolding occurs at lower forces, increasing the average dissipation, $\langle W_d\rangle \geq 0$. The Crooks FT and the Jarzynski equality, that is Eq.\eqref{eq.Jarz} with $\gamma = 1$ and Eq.\eqref{eq.FT-1stTF} with $\Upsilon_M = 0$, are fulfilled, because, no feedback is involved. 
\subsection{Mean-field approximation (MFA)}\label{sec:JMA}
Neglecting multiple hopping transitions between F and U along $\rightarrow$ ($\leftarrow$) implies that, once the molecule jumps to U (F) at a given $\lambda$ it remains in that state until reaching $\lambda_{max}$ ($\lambda_{min}$). Therefore we identify $\psi \left( \lambda \right) = c \cdot p_\rightarrow ^F \left(\lambda,r_F \right) p_\rightarrow ^U\left( \lambda,r_F\right)$ and $\widetilde{\psi} \left( \lambda \right) = c' \cdot p_\leftarrow ^F\left( \lambda,r_F \right) p_\leftarrow ^U\left( \lambda,r_F \right)$ where $\psi$ (${\widetilde{\psi}}$) is the normalized probability along $\rightarrow$ ($\leftarrow$) to observe U at a given value of $\lambda$ for the first (last) time at the single loading (unloading) rate $r_F$ and $p^{\sigma}_{\rightarrow(\leftarrow)}(\lambda,r)$ is the fraction of trajectories observed in $\sigma (=F,U)$ at $\lambda$ with loading rate $r$. $c$ and $c'$ are normalizing factors. Terms $J\left(\lambda\right)$, $\Upsilon_\infty$ in Eqs.(\ref{eq.dFT-CTF}b,\ref{eq.FT-CTF}b) give,
\begin{subequations}\label{eq.JMA}
\begin{align}
J_{MFA}\left(\lambda\right) = I_F\left(\lambda\right)+I_U\left(\lambda\right) + \log\left(\frac{c'}{c}\right)\ \\
\Upsilon_{MFA} = \log{\left(\int_{\lambda_{min}}^{\lambda_{max}}{\psi \left( \lambda \right)}e^{J_{MFA}(\lambda)}\ \ d\lambda\right)}\ = \nonumber \\
= \log \left( \frac{\int_{\lambda_{min}}^{\lambda_{max}}{p_\leftarrow^F\left(\lambda,r_F\right)p_\leftarrow^U\left(\lambda,r_U\right)}d\lambda}{\int_{\lambda_{min}}^{\lambda_{max}}{p_\leftarrow^F(\lambda,r_F)p_\leftarrow^U(\lambda,r_F)}d\lambda}\right)
\end{align}
\end{subequations}
with $I_F\left(\lambda\right)$ and $I_U\left(\lambda\right)$ in Eq.(\ref{eq.JMA}a) the partial thermodynamic information as in Eq.\eqref{eq.dFT-DTF} for DTF at a given $\lambda$. For the experiments of hairpin L4 the transition state is located at half-distance of the molecular extension, resulting into nearly symmetric forward and reverse processes without feedback. Therefore, $c'\cong c$ and $\log(c'/c)\cong 0$. Equation (\ref{eq.JMA}a) shows $J_{MFA} \left( \lambda \right)$ in CTF equals the sum of the $I_\sigma \left( \lambda \right)$ corresponding to DTF. $\Upsilon_{MFA}$ in Eq.(\ref{eq.JMA}b) is a good approximation for $\Upsilon_\infty$ under highly irreversible pulling conditions where the molecule executes a single unfolding (folding) transition during $\rightarrow$ ($\leftarrow$). Equations (\ref{eq.JMA}a,\ref{eq.JMA}b) can be approximated using the Bell-Evans model for $p_{\rightarrow(\leftarrow)} ^\sigma\left( \lambda,r \right)$ as shown in Figures \ref{fig2}c,\ref{fig4}c.
\section{\label{seq.Results}Results}
\subsection{First-time feedback (1$^{\bf st}$TF).}
\label{sec:1stTF}
% 
% _________________ TABLE 1 ________________
\renewcommand{\arraystretch}{2}
\begin{table*} 
    \centering
    \begin{tabular}{|c||c|c|c|c|c|c|}
    \hline
          \textbf{protocol} & \textbf{(a)} & \textbf{(b)} & \textbf{(c)} & \textbf{(d)} & \textbf{(e)} & \textbf{(f)}    \\\hline \hline
        
    1$^{\rm st}$TF & $p_{\rightarrow,k}^\sigma\left(r\right)$ & $p_{\leftarrow,k}^\sigma\left(r\right)$ & $\psi_k$ & ${\widetilde{\psi}}_k$ & $J_k$ & $\Upsilon_M$ \\ \hline
    
    DTF & $p_{\rightarrow}^\sigma\left(r_F\right)$ & $p_{\leftarrow}^\sigma\left(r_U\right)$ & \begin{tabular}[c]{@{}l@{}} $\psi_1=p_\rightarrow^U\left(r_F\right)$ \\ $\psi_2=p_\rightarrow^F\left(r_F\right)$ \end{tabular} & \begin{tabular}[c]{@{}l@{}} ${\widetilde{\psi}}_1=p_\leftarrow^U\left(r_F\right)$ \\ ${\widetilde{\psi}}_2=p_\leftarrow^F\left(r_F\right)$ \end{tabular} & \begin{tabular}[c]{@{}l@{}} ${\ \ J}_1=I_U$ \\ ${\ \ J}_2=I_F$ \end{tabular} & $\Upsilon_2$   \\ \hline
    
    CTF & $p_{\rightarrow}^U\left(r_F\right)$ & \begin{tabular}[c]{@{}l@{}} $p_\leftarrow^U\left(\lambda,r_U\right)$ \\ $p_\gets^U\left(\lambda,r_F\right)$ \end{tabular} & $\psi(\lambda)$ & $\widetilde{\psi}(\lambda)$ & $J(\lambda)$ & $\Upsilon_\infty$ \\ \hline
        \end{tabular}
\caption{\textbf{Probabilities and densities in the different protocols}: Probabilities (a,b,c,d) must be measured to calculate the corresponding densities $\rho_\rightarrow \left( W|k \right)$, $\rho_ \leftarrow \left( -W|k \right)$, verify the feedback-FT and extract partial and full thermodynamic information $J$, $\Upsilon$ (columns e,f). All quantities are properly normalized in 1$^{\rm st}$TF:  $\sum_{\sigma} {p_{\rightarrow,k} ^\sigma\left(r \right)} = \sum_{\sigma}{p_{\leftarrow,k}^\sigma\left(r\right)} = \int{ dW \rho_\rightarrow \left (W|k \right)} = \int{dW \rho_ \leftarrow \left( -W|k \right)} = 1$ for all $k$ and $\sum_{k=1} ^{M} \psi_k = \sum_{k=1} ^{M} {\widetilde{\psi}}_k = 1$; $\int{ dW \rho_ \rightarrow \left( W \right)} = \int{ dW \rho_\leftarrow \left( -W \right)} = 1$. Distributions in CTF are also normalized: $\int{p_{\rightarrow,\leftarrow}^U\left(\lambda,r\right) d\lambda} = \int \psi d\lambda = \int{\widetilde{\psi} d\lambda} = 1;\ \int { dW \rho_{\rightarrow,\leftarrow}} = 1$}
\label{Tab.1} 
\end{table*}
To address DTF and CTF in full generality we introduce the 1$^{\rm st}$TF protocol, a repeated-time feedback protocol suitable for pulling experiments (Fig.\ref{fig1}d) that interpolates between DTF and CTF. In the 1$^{\rm st}$TF protocol the trap-position range $[\lambda_{min}, \lambda_{max}]$ is discretized in $M+1$ steps, $\left\{\lambda_k;0 \le k\le M\right\}$ with boundaries $\lambda_0=\lambda_{min} ; \lambda_M=\lambda_{max}$. The feedback protocol needs at least one intermediate measure position, so $M \geq 2$. The initially folded (F) molecule is pulled at a loading rate $r_F$ and measurements are taken at every discrete position $\lambda_k$ along the forward ($\rightarrow$) process (Fig.\ref{fig1}d). The state $\sigma_k (=F,U)$ is monitored at every $\lambda_k$ until a $\lambda_{k^\ast}$ is reached where the molecule is observed to be in U for the first time $\left\{{\sigma_{k^\ast}=U;\ \sigma}_k=F\ ,\ 0\le k<k^\ast\right\}$. At $\lambda_{k^\ast}$ the loading rate is changed to $r_U$ (Fig.\ref{fig1}d, top). Like in generic isothermal feedback processes the conditional reverse ($\leftarrow$) process is the time-reverse of $\rightarrow$ \cite{Sag2014}: the unloading rate is equal to $r_U$ from $\lambda_{max}$ to $\lambda_{k^\ast}$, after which the unloading rate is changed back to $r_F$ between $\lambda_{k^\ast}$ and $\lambda_{min}$ (Fig.\ref{fig1}d, bottom). Therefore, no feedback is implemented on $\leftarrow$. Let $p_{\rightarrow,k}^\sigma\left(r\right)\ (p_{\leftarrow,k}^\sigma\left(r\right))$ be the probability to observe the molecule in state $\sigma$ (F or U) at $\lambda_k$ along $\rightarrow(\leftarrow)$ at the pulling rate $r$. We derived a detailed and full work-FT for 1$^{\rm st}$TF (Appendix \ref{sec:1stTF_der}). The latter reads,
\begin{eqnarray}
\label{eq.FT-1stTF}
    & & \frac{\rho_{\rightarrow}(W)}{\rho_{\leftarrow}(-W)} = \exp\left[ \frac{W-\Delta G_{FU}+k_BT\Upsilon_M}{k_BT}\right] ~ \nonumber \\
    & &  {\rm with }~ \Upsilon_M = \log \left( \sum_{k=1}^M\frac{p_{\leftarrow,k}^U(r_U)}{p_{\leftarrow,k}^U(r_F)}\widetilde{\psi}_k \right) .
\end{eqnarray}
$W$ is the work measured between $\lambda_{min}$ and $\lambda_{max}$ (Methods \ref{sec:pullexp}) while $\Delta G_{FU}$ is the free energy difference between the state U at $\lambda_{max}$ and the state F at $\lambda_{min}$, $\Delta G_{FU} = G_U\left( \lambda_{max} \right) - G_F\left( \lambda_{min} \right)$. Work distributions are given by,
\begin{eqnarray}
& & \rho_\rightarrow\left( W \right) = \sum_{k=1} ^{M}{ \rho_\rightarrow \left(W|k \right)\psi_k}\label{eq.rhoFFTF}\\
& & \rho_\leftarrow \left( -W \right) = e^{-\Upsilon_M} \sum_{k=1} ^{M} {\rho_\leftarrow \left( -W|k \right)} \frac{p_{\leftarrow,k} ^U \left(r_U \right)} {p_{\leftarrow,k} ^U\left( r_F \right)} {\widetilde{\psi}}_k\label{eq.rhoRFTF}.
\end{eqnarray}
Here $\rho_\rightarrow\left(W|k\right)$ and $\rho_\leftarrow\left(-W|k\right)$ are the partial work distributions along $\rightarrow$ and $\leftarrow$ conditioned to those $\rightarrow$ paths where U is observed for the first time at $\lambda_k$. $\psi_k$ (${\widetilde{\psi}}_k$) $\left(1 \le k \le M\right)$ is the probability along $\rightarrow$ ($\leftarrow$) to observe U at $\lambda_k$ for the first (last) time at the single loading (unloading) rate $r_F$. Note that while $\psi_k$ can be measured from $\rightarrow$ pulls with the feedback on, the \textit{reverse} quantities $\widetilde{\psi}_k$, $p_{\leftarrow,k}^U\left(r_F\right)$, and $p_{\leftarrow,k}^U\left(r_U\right)$ are measured from reverse pulls at either the unloading rates $r_F$ or $r_U$ without feedback. Table \ref{Tab.1} summarizes these definitions. $\Upsilon_M$ in Eq.\eqref{eq.FT-1stTF} denotes the thermodynamic information, 
\begin{eqnarray}\label{eq.Ym-1stTF}
    \Upsilon_M=\log{\left(\sum_{k=1}^{M}{\psi_k\exp{J_k}}\right)}\ ~ \nonumber \\
    {\rm with}~ J_k=\log{\left(\frac{p_{\leftarrow,k}^U\left(r_U\right){\widetilde{\psi}}_k}{p_{\leftarrow,k}^U\left(r_F\right)\psi_k}\right)},
\end{eqnarray}
%
% ///////////////////////////////////////////////
% _____________________________________ FIGURE 2 
\begin{figure*}
\centering\includegraphics{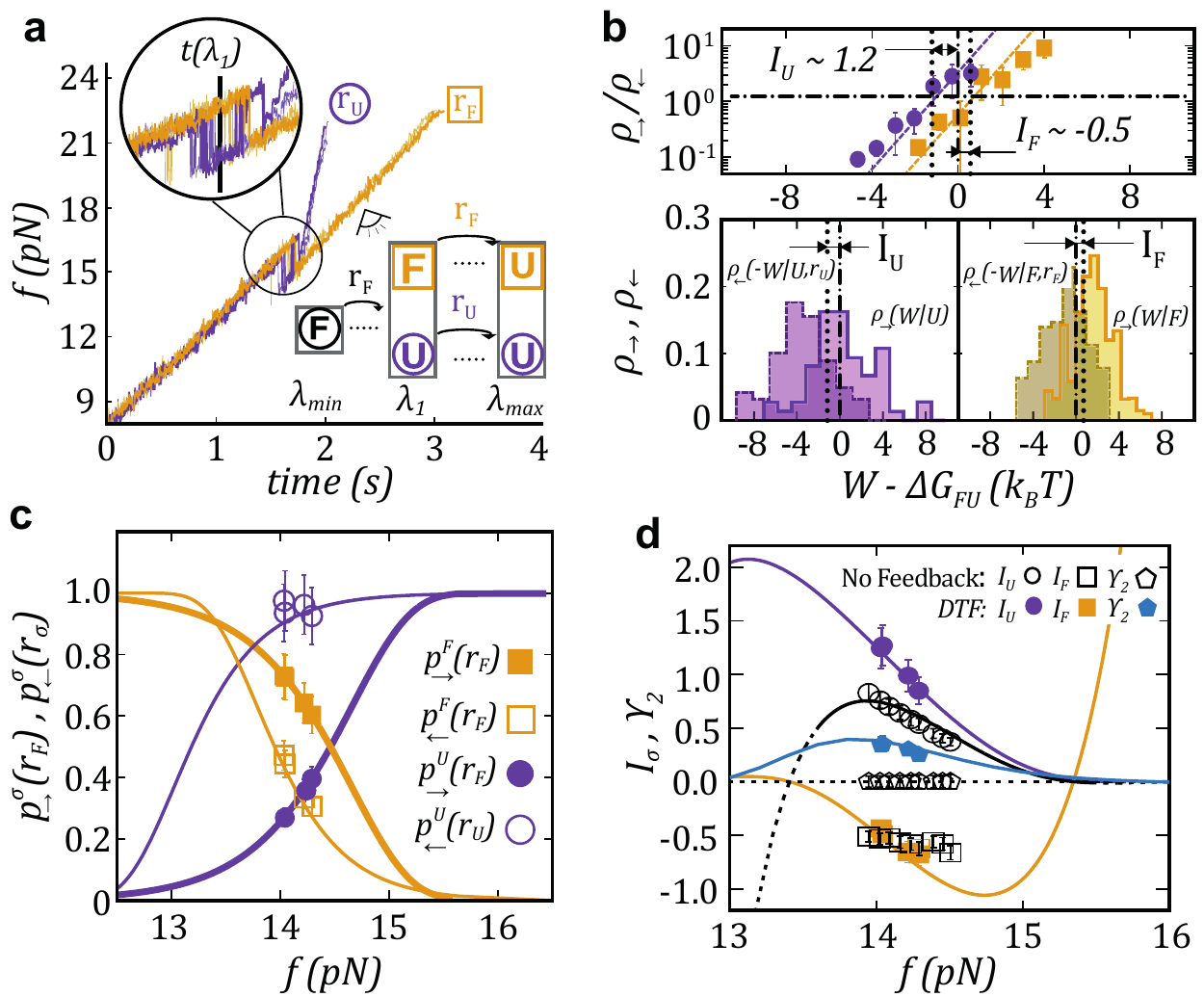}
    \caption{{\bf Discrete-Time Feedback (DTF).} (a) Experimental force-time unfolding curves. The molecule is pulled at $r_F=4$pN/s and the observation is made at $\lambda_1$: if the molecule is folded ($\sigma = F$) the pulling rate remains unchanged ($r=r_F$, orange curves); if it is unfolded ($\sigma = U$) the pulling rate is changed to $r_U = 17$pN/s, making refolding events less likely (purple curves). The schematic summarizes the feedback protocol and the reaction to the measurement outcome. (b) Bottom: Detailed feedback-FT: $\rho_\rightarrow\left(W|\sigma\right)$ (solid lines) and $\rho_\leftarrow\left(-W|\sigma,r_\sigma\right)$ (dashed lines) for $\sigma = F, U$ (orange, purple) trajectories. Top: Test of the detailed feedback-FT Eq.\eqref{eq.dFT-DTF} for $\sigma = F,U$. (c) Probabilities $p_\rightarrow^\sigma\left(r_F\right)$, $p_\leftarrow^\sigma \left(r_\sigma\right)$ and (d) information terms $I_\sigma$ and $\Upsilon_2$ as a function of the force in U measured at $\lambda_1$. Orange (purple) data are for $\sigma = F(U)$. In (d) we also show $I_\sigma$ without feedback (empty symbols). Theoretical predictions from the Bell-Evans model are shown as solid lines in (c,d). }
    \label{fig2}
\end{figure*}
% ____________________________________________________________________
where $J_k$ is the \textit{partial} thermodynamic information along $\rightarrow$, restricted to those paths where U is observed for the first time at $\lambda_k$. Equation \eqref{eq.Ym-1stTF} expresses $\Upsilon_M$ as a partition sum or potential of mean force over the partial contributions $J_k$. Note that for $r_U = r_F$, $J_k = \log{\left( {{\widetilde{\psi}}_k} / {\psi_k} \right)}$ and $\Upsilon_M = \log{\left( \sum_{k=1} ^{M} {\widetilde{\psi}}_k \right)}=0$. In this case feedback is without effect and Crooks FT \cite{Cro1999} is recovered. The 1$^{\rm st}$TF protocol has DTF and CTF as particular cases: DTF corresponds to $M = 2$; whereas $M\rightarrow\infty$ yields CTF.

Let us assume that the time between consecutive measurements at $\lambda_k$ is equal to $\tau$, in which case $M\rightarrow\infty$ corresponds to $\tau\rightarrow 0$. To implement CTF experimentally we take the lowest possible value of $\tau$ as dictated by the maximum data acquisition frequency of the instrument ($\tau\approx1$ms). Below we report feedback experiments in DTF and CTF and measure $\Upsilon$, $\eta_I$ and $\eta_M$ in the regime $r_U > r_F$.
%________________________________________________
\subsection{Discrete-time feedback (DTF, M = 2)}
\label{sec:DTF}
In DTF the pulling rate along $\rightarrow$ is changed from $r_F$ to $r_U$ at a given trap position $\lambda_1$ if $\sigma = U$, otherwise it remains unchanged (Figs.\ref{fig1}c, \ref{fig2}a). Therefore if $\sigma = F$ at $\lambda_1$ the pulling rate is constant and equal to $r_F$ throughout the pulling cycle. Whereas if $\sigma = U$ at $\lambda_1$ the pulling rate along $\leftarrow$ starts at $r = r_U$ and switches back to $r_F$ at $\lambda_1$. A detailed feedback-FT can be derived either from the detailed form of Eq.\eqref{eq.FT-1stTF} for $M = 2$ or from the extended fluctuation relation \cite{Juni2009} (Table \ref{Tab.1} and Appendix B)
\begin{eqnarray}\label{eq.dFT-DTF}
    & &\frac{\ \rho_\rightarrow\left(W|\sigma\right)}{\rho_\leftarrow\left(-W|\sigma,r_\sigma\right)}=\exp{\left[\frac{W-\mathrm{\Delta}G_{FU}+k_BT\ I_\sigma}{k_BT}\right]}\ \nonumber \\
    & &{\rm with} ~ I_\sigma\equiv\log{\left(\frac{p_\leftarrow^\sigma\left(r_\sigma\right)}{p_\rightarrow^\sigma\left(r_F\right)}\right)}\,;\,\sigma=\left\{F,U\right\}\ 
\end{eqnarray}
where $\sigma \left(=F,U\right)$ is the measurement outcome at $\lambda_1$ along $\rightarrow$. $\rho_\rightarrow\left(W|\sigma\right)$ and $\rho_\leftarrow \left( W|\sigma, r_\sigma \right)$ are the (normalized) work distributions conditioned to those trajectories passing through $\sigma$ at $\lambda_1$ along $\rightarrow$ and  $\leftarrow$, respectively. The $p_\rightarrow ^\sigma\left( r_F \right)$, $p_\leftarrow ^\sigma\left( r_\sigma \right)$ are the probabilities to measure $\sigma$ at $\lambda_1$ along $\rightarrow$ and $\leftarrow$, at the respective pulling rates. Finally, $I_\sigma$ is the partial thermodynamic information of measurement outcome $\sigma$. Note that $I_\sigma$ can take any sign depending on the ratio ${p_\leftarrow^\sigma\left(r_\sigma\right)}/{p_\rightarrow^\sigma\left(r_F\right)}$, which can be larger or smaller than $1$. Moreover, while all trajectories in $\rightarrow$ are classified in one of the two groups $\sigma=\left\{F,U\right\}$, only those that revisit again the same $\sigma$ contribute to $\rho_\leftarrow\left(-W|\sigma,r_\sigma\right)$. Therefore, the normalization condition along $\rightarrow$, $\sum_{\sigma} {p_\rightarrow ^\sigma\left( r_F \right)} = 1$, is not applicable to $\leftarrow$, i.e., $\sum_{\sigma} {p_\leftarrow^ \sigma\left( r_\sigma \right)} \neq 1$. 
% ///////////////////////////////////////////////
% _____________________________________ FIGURE 3 
\begin{figure*}
\centering\includegraphics{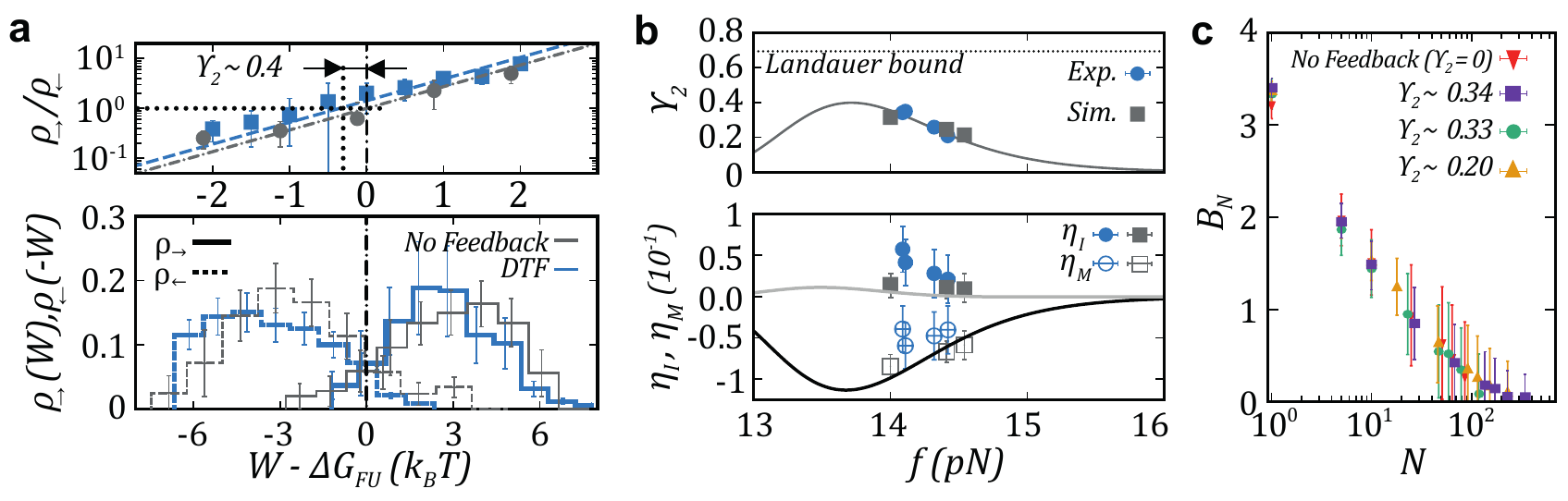}
    \caption{{{\bf DTF efficiencies $\eta_I,\eta_M$}.} (a) Top: Test of the full feedback-FT Eq.(\ref{eq.FT-DTF}a) with shift $\Upsilon_2$ (blue) with respect to the No Feedback case (gray). Bottom: Work distributions (b) Thermodynamic information, $\Upsilon_2$ (top), and efficiencies $\eta_I$ and $\eta_M$ (bottom) are shown at different measurement positions (grey: simulations; blue: experiments). Solid lines are the Bell-Evans model in the single-hopping approximation (Appendix \ref{sec:BellEvans}). In general, $\Upsilon_2$ is bounded from above by the Landauer limit for binary measurements ($\log 2$, dashed line). (c) Bias for three different DTF conditions compared to the non-feedback case. There is no improvement in free energy prediction.}
    \label{fig3}
\end{figure*}
% ____________________________________________________________________
Equation \eqref{eq.dFT-DTF} for $\sigma = \left\{F,U \right\}$ permits to extract $\Delta G_{FU}$ from DTF experiments. First, we apply the protocol without feedback ($r_U = r_F$) as a consistency check (Fig.\ref{fig1}b). We find that Crooks FT \cite{Cro1999} is satisfied with work distributions ($\rightarrow$, $\leftarrow$) crossing at a $\Delta G_{FU}$ consistent with bulk predictions, (S1, Supp. Info.). Next, we apply DTF with $r_U > r_F$ to extract partial work distributions $\rho_\rightarrow\left( W|\sigma \right)$ and $\rho_\leftarrow\left( -W| \sigma, r_\sigma \right)$ by classifying trajectories depending on the outcome $\sigma$ at $\lambda_1$ and the protocol under which they are operated. Figure \ref{fig2}b (bottom) shows results for $\sigma = F,U$ for $r_F = 4$ pN/s,  $r_U = 17$ pN/s. For the F (U) subset we find that the work distributions are shifted rightwards (leftwards) with respect to the non-feedback case with crossing points ($W_\sigma^\ast$) such that $W_F^\ast>\Delta G_{FU}$ ($W_U^\ast< \Delta G_{FU}$). These shifts reflect the fact that hairpin unfolding is on average more (less) energy-costly for the F (U) subset than without feedback at the pulling rate $r_F$. From Eq.\eqref{eq.dFT-DTF} the measured shift, defined as $W_ \sigma ^\ast - {\Delta}G_{FU}$, equals $-k_BT I_\sigma$. The $\rho_ \rightarrow \left( W|\sigma \right)$ and $\rho_ \leftarrow \left( -W|\sigma,r_\sigma \right)$ fulfil Eq.\eqref{eq.dFT-DTF} crossing at values $W_\sigma^\ast = \Delta G_{FU} - k_BTI_\sigma$ with $I_F\approx -0.5$ and $I_U \approx 1.20$ (Fig.\ref{fig2}b, top). In Figure \ref{fig2}c,d we show $p_\rightarrow ^\sigma \left( r_F \right)$, $p_\leftarrow ^\sigma \left( r_\sigma \right)$ and $I_{F,U}$ versus the force in U at $\lambda_1$ together with predictions based on the Bell-Evans model (Appendix \ref{sec:BellEvans}). We choose force as a reference value to present the results. Force is more informative than the trap position $\lambda$, the latter being the relative distance between the trap position and an arbitrary initial position in the light-lever detector.

Combining the detailed feedback-FT Eq.\eqref{eq.dFT-DTF} for $\sigma = F,U$ yields to the full work-FT Eq.\eqref{eq.FT-1stTF} for $M = 2$ (Appendix B),
\begin{subequations}\label{eq.FT-DTF}
\begin{align}
        \frac{\rho_\rightarrow(W)}{\rho_\leftarrow(-W)}=\exp{\left[\frac{W-{\Delta}G_{FU}+k_BT  \Upsilon_2}{k_BT}\right]\  } \\
          {\rm with}\ ~ \Upsilon_2 = \log{\left(\sum_{\sigma=F,U}{p_\leftarrow^\sigma\left(r_\sigma\right)}\right)} = \nonumber \\
          =\log{\left(\sum_{\sigma=F,U}{p_\rightarrow^\sigma\left(r_F\right)}\exp{I_\sigma}\right)},
\end{align}
\end{subequations}
$\Upsilon_2$ being the thermodynamic information. The forward and reverse work distributions are given by, 
\begin{eqnarray}
\rho_\rightarrow\left(W\right) = \sum_{\sigma} {p_\rightarrow^ \sigma\left( r_F \right) \rho_\rightarrow \left( W|\sigma \right)}\label{eq.rhoFDTF}\\  \rho_ \leftarrow \left( -W \right) = e^{-\Upsilon_2}\sum_{\sigma} {p_ \leftarrow ^\sigma \left( r_\sigma \right) \rho_ \leftarrow \left( -W|\sigma, r_\sigma \right)} \label{eq.rhoRDTF} 
\end{eqnarray}
For $r_F = r_U$ we have $p_\leftarrow^F \left( r \right) + p_\leftarrow^U \left (r \right) = 1$ yielding $\Upsilon_2 = 0$ and Crooks FT \cite{Cro1999} as expected. Figure \ref{fig3}a tests Eq.(\ref{eq.FT-DTF}a) and Figure \ref{fig3}b shows $\Upsilon_2$ and the efficiencies $\eta_I$, $\eta_M$  obtained in experiments. Results are compared with numerical simulations of the DNA pulling experiments (see details in S2 of Supp. Info.) and a prediction by the Bell-Evans model (Appendix \ref{sec:BellEvans}). For $r_U > r_F$, $p_ \leftarrow ^F \left( r_F \right) + p_\leftarrow ^U \left( r_U \right) > 1$ and $\Upsilon_2 > 0$, whereas ${\Upsilon}_2 < 0$ for $r_U < r_F$. In general, $-\infty < \Upsilon_2 \le \log{2}\ (0\le p_\leftarrow ^\sigma \left( r_\sigma \right) \le 1)$ showing that the Landauer bound holds for two-state molecules pulled under DTF. Saturating the bound, $\Upsilon_2 = \log2$, requires full reversibility \cite{Horo2011}, i.e., $p_ \leftarrow ^\sigma \left( r_\sigma \right) = 1$, which is obtained for arbitrary $r_F$ in the limit $r_U \rightarrow \infty$ ($p_\leftarrow ^U \left( r_U \right) = 1)$ and $\lambda_1 \rightarrow \lambda_{min}$ (i.e., maximally stable F or  $p_\leftarrow ^F \left( r_F \right) = 1$). We find $\eta_M \approx -0.05 < \eta_I \approx 0.04$ showing that information-to-measurement conversion is much less efficient than information-to-work conversion. Moreover, $\eta_M<0$ throughout the whole force range shows that DTF does not improve free energy prediction.

To better understand this result we have calculated the efficiencies $\eta_I$ and $\eta_M$ for DTF in the two-states Bell-Evans model using the single-hopping approximation (Appendix \ref{sec:BellEvans}). Figure \ref{fig3}b shows that the analytical results capture the trend of the experimental data but systematically underestimate the measured efficiencies $\eta_I$ and $\eta_M$. As explained in Appendix \ref{sec:BellEvans} the single-hopping approximation neglects multiple transitions after the measurement position at $\lambda_1$. Therefore, the analytical results derived in Eqs.(\ref{Wd_BE},\ref{Af_BE},\ref{AWd_BE}) in Appendix \ref{sec:BellEvans} are lower bounds to the true efficiencies. The fact that $\eta_M<0$ throughout the force range shows that although DTF does reduce dissipation it does not improve free energy prediction. This conclusion is supported by the results shown in Fig.\ref{fig3}c. There we plot the experimental free energy bias Eq.\eqref{eqBias} as a function of the number of pulling experiments $N$ at the conditions shown in panel b: bias with feedback does not decrease with respect to the non-feedback case (downward pointing red triangles).

%//////////////////////////////////////////////////////////////////
\subsection{Continuous-time feedback (CTF, M$\to \infty$)}
\label{sec:CTF}
%
% _____________________________________ FIGURE 4
\begin{figure*}
\centering\includegraphics{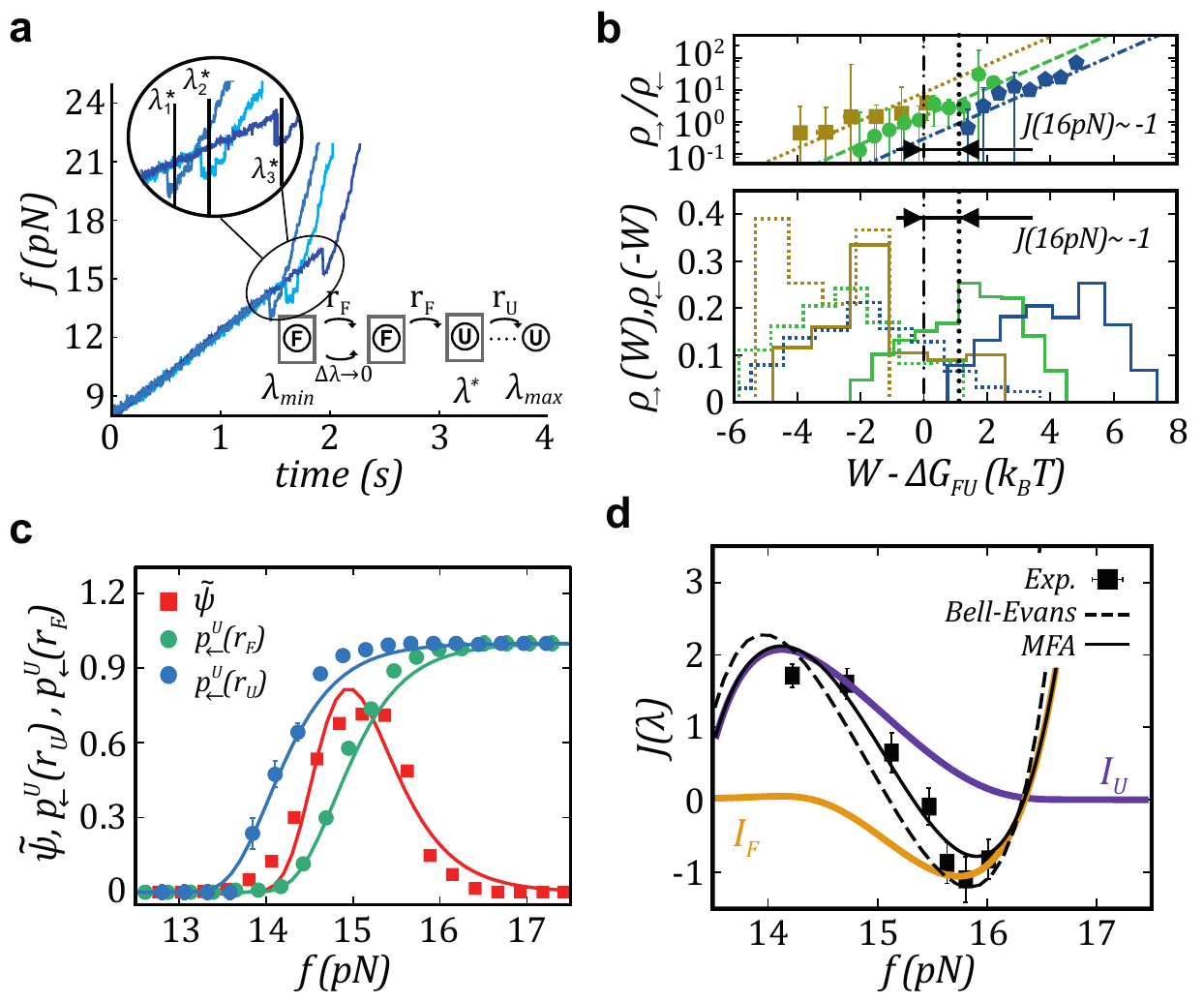}
    \caption{{\bf Continuous-time feedback (CTF).} (a) Experimental unfolding curves. The molecule is pulled at $r_F$=5pN/s along $\rightarrow$. Upon detection of the first unfolding event the pulling rate is changed to $r_U$=23pN/s. (b) Experimental test of the detailed feedback-FT Eq.(\ref{eq.dFT-CTF}a) and measurement of $J\left(\lambda\right)$. FDCs have been classified according to the value of $\lambda$ at which the earliest unfolding event is detected. FDCs have been grouped in bins of equal width $\Delta\lambda\sim 6$nm and Eq.({\ref{eq.dFT-CTF}}a) applied to each interval. We show work histograms for three different bins. The value of $J\left(\lambda\right)$ for the bin corresponding to $f$=16pN is highlighted. We use force due to its one-to-one correspondence along the folded branch with $\lambda$. (c) $p_\leftarrow^U\left(\lambda,r_U\right)$, $p_\leftarrow^U\left(\lambda,r_F\right)$ and last-folding density $\widetilde{\psi}(\lambda)$. Solid lines correspond to fits to the Bell-Evans model. (d) The $J\left(\lambda\right)$ from the detailed feedback-FT Eq.(\ref{eq.dFT-CTF}a) (squares) obtained from the data in panel (b) compared to the theoretical prediction by the Bell-Evans model (c) applying Eq.(\ref{eq.dFT-CTF}b) (dashed line). Also, the MFA approximation Eq.(\ref{eq.JMA}a) with $\log(c'/c)\sim0$ is shown as a solid line. $I_F\left(\lambda\right)$ (orange) and $I_U\left(\lambda\right)$ (purple) are from Eq.\eqref{eq.dFT-DTF} using the Bell-Evans model. The values of  $p_\leftarrow^U(r_U)$, $p_\leftarrow^U(r_F)$, $\widetilde{\psi}$, $J$, $I_F$, $I_U$ in (c,d) are plotted versus $f(\lambda)$. }
    \label{fig4}
\end{figure*}
% ____________________________________________________________________

CTF (Figs.\ref{fig1}c, \ref{fig4}a) is obtained from Eqs.(\ref{eq.FT-1stTF},\ref{eq.Ym-1stTF}) in the limit of $\Delta \lambda(= \lambda_{k+1} - \lambda_k)$, $\tau \rightarrow 0$. The detailed feedback-FT reads (Table \ref{Tab.1} and Appendix \ref{sec:1stTF_der}),
% ///////////////////////////////////
% ___________________________________ Figure 5
\begin{figure*}
\centering\includegraphics{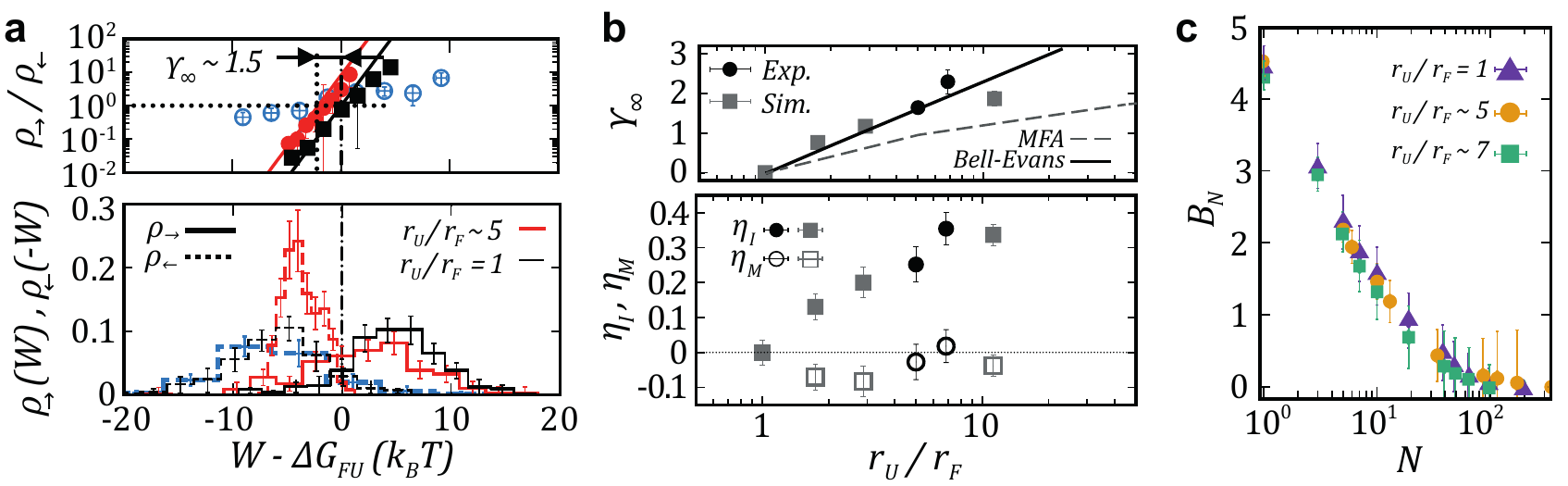}
    \caption{{\bf CTF efficiencies $\eta_I,\eta_M$.} (a) Top: Test of the full feedback-FT Eq.(\ref{eq.FT-CTF}a) (red circles) with shift $\Upsilon_\infty$ with respect to the non-feedback case (black squares). Blue circles show a negative test with an unweighed reverse work distribution (see text). Bottom: Work distributions. (b) (Top) Thermodynamic information $\Upsilon_\infty$ and (Bottom) efficiencies $\eta_I$ and $\eta_M$ from the experimental data (circles) and simulations (squares) versus $r_U/r_F$. We also show the theoretical predictions for $\Upsilon_\infty$, Eq.(\ref{eq.FT-CTF}b), using $p_\leftarrow^U(r_U)$, $p_\leftarrow^U(r_F)$, $\widetilde{\psi}$ from Fig.\ref{fig4}c (continuous lines) and the MFA Eq.(\ref{eq.JMA}b) (dashed line). Notice that $\Upsilon_\infty$ is unbounded from above and $\eta_M < 0$. (c) Bias for the studied molecules with CTF (squares, circle) compared with the non-feedback case. There is no improvement in free energy prediction.} 
    \label{fig5}
\end{figure*}
% ____________________________________________________________________
%
\begin{subequations}\label{eq.dFT-CTF}
\begin{align}
    \frac{\rho_\rightarrow\left(W|\lambda\right)}{\rho_\leftarrow\left(-W|\lambda\right)}=\exp{\left[\frac{W-{\Delta}G_{FU}+k_BT\ J(\lambda)}{k_BT}\right]}; \\  J(\lambda)={\log{\left(\frac{p_\leftarrow^U\left(\lambda,r_U\right)\widetilde{\psi}\left(\lambda\right)}{p_\leftarrow^U\left(\lambda,r_F\right)\psi(\lambda)}\right)}}
\end{align}
\end{subequations}
while the full feedback-FT reads,
\begin{subequations}\label{eq.FT-CTF}
\begin{align}
    \frac{\rho_\rightarrow(W)}{\rho_\leftarrow(-W)}=\exp{\left[\frac{W-{\Delta}G_{FU}+k_BT\Upsilon_\infty}{k_BT}\right]} ; \\  \Upsilon_\infty=\log\ \left(\int_{\lambda_{min}}^{\lambda_{max}}\frac{p_\leftarrow^U\left(\lambda,r_U\right)}{p_\leftarrow^U\left(\lambda,r_F\right)}\widetilde{\psi}(\lambda)d\lambda\right) = \nonumber \\ =\log{\left(\int_{\lambda_{min}}^{\lambda_{max}}{\psi\left(\lambda\right)\exp{J(\lambda)}}d\lambda \right)}
\end{align}
\end{subequations}
with the forward and reverse work distributions given by, 
\begin{eqnarray}
\rho_\rightarrow \left( W \right) = \int_{ \lambda_{min} } ^{ \lambda_{max} } {\rho_\rightarrow \left( W|\lambda \right)} \psi(\lambda) d\lambda\label{eq.rhoFCTF}\\
\rho_\leftarrow(-W) = e^{-\Upsilon_\infty} \int_{ \lambda_{min} } ^{ \lambda_{max} } {\rho_\leftarrow \left( -W|\lambda \right)} \cdot\nonumber\\
\frac{ p_\leftarrow ^U\left(\lambda,r_U\right) } { p_\leftarrow ^U \left( \lambda,r_F \right)} \widetilde{\psi} \left( \lambda \right) d\lambda\label{eq.rhoRCTF}
\end{eqnarray}
where $\psi(\lambda)$ ($\widetilde{\psi}(\lambda)$) is the probability density to observe the first (last) unfolding (folding) event $F\rightarrow U$ ($F\leftarrow U$) along $\rightarrow$ ($\leftarrow$); $p_\leftarrow^U\left(\lambda,r\right)$ is the probability density of the molecule being in U at $\lambda$ along $\leftarrow$ at the unloading rate $r$. Similarly to $I_\sigma$ in Eq.\eqref{eq.dFT-DTF}, if we define $I_\sigma ^{rr\prime} \left( \lambda \right) = \log{\left( \frac{p_\leftarrow ^\sigma\left( \lambda, r^\prime \right)} {p_\rightarrow ^\sigma \left(\lambda, r\right)} \right)}$, we have $J\left( \lambda \right) = I_U^{r_F r_U}\left(\lambda\right) - I_U ^{r_F r_F} \left( \lambda \right) + I_\psi \left(\lambda\right)$ with $I_\psi\left(\lambda\right)=\log{\left(\frac{\widetilde{\psi}\left(\lambda\right)}{\psi(\lambda)}\right)}$. Notice that for $r_U=r_F$ (no feedback), $\rho_ \leftarrow \left( -W | \lambda \right) = \rho_\leftarrow \left( -W \right)$ and ${\Upsilon}_\infty = 0$, but $J\left( \lambda \right) = I_\psi \left( \lambda \right) \neq 0$. Equations (\ref{eq.dFT-CTF}b, \ref{eq.FT-CTF}b) can be further simplified by neglecting multiple hopping transitions between F and U. In this mean-field approximation (MFA), $J\left(\lambda\right)$ interpolates the $I_\sigma$ in Eq.\eqref{eq.dFT-DTF} and $\Upsilon_\infty$ only depends on the $p_\leftarrow\left(\lambda,r\right)$ (Methods \ref{sec:JMA}). 

We tested CTF in DNA hairpin pulling experiments (Fig.\ref{fig4}a). The molecule initially in F is pulled from $f_{min}=8$pN at $r_F=5$pN/s and the state monitored by recording the force every $\tau=1$ms until the first force jump is observed at a given trap position $\lambda^\ast$. The force rip corresponds to the unzipping of the 44 nucleotides of the DNA hairpin and indicates that state U has been visited for the first time at $\lambda^\ast$. Then the pulling rate is increased to $r_U=23$pN/s until the maximum force is reached, $f_{max}=22$pN. For the reverse process the optical trap moves backwards at $r_U= 23$pN/s until $\lambda^\ast$ is reached and the pulling rate switched back to $r_F= 5$pN/s. By repeatedly pulling we collect enough statistics to test Eqs.(\ref{eq.dFT-CTF}a, \ref{eq.FT-CTF}a) and measure $J\left(\lambda\right)$ and ${\Upsilon}_\infty$. In Figure \ref{fig4}b (bottom) we plot $\rho_\rightarrow \left( W| \lambda \right)$, $\rho_\leftarrow \left( -W| \lambda \right)$ for three selected $\lambda^\ast$ while in the top panel we test Eq.(\ref{eq.dFT-CTF}a). By determining the crossing work values between $\rho_\rightarrow$ and $\rho_\leftarrow$, $W^\ast \left(\lambda \right) = {\Delta}G_{FU} - k_BTJ\left(\lambda\right)$, we extract $J\left(\lambda\right)$.  

Figure \ref{fig4}c shows the values of $\widetilde{\psi}$ and $p_\leftarrow^U$ directly determined from experimental FDCs for the two loading rates, $r_F = 5$pN/s and $r_U = 23$pN/s (symbols) as a function of force. This has been fitted to the Bell-Evans model (solid lines) to extract the kinetic parameters of hairpin L4, useful to compare with the simulations. Figure \ref{fig4}d shows the experimental values of $J\left(\lambda\right)$ determined from the detailed feedback-FT Eq.(\ref{eq.dFT-CTF}a) (filled squares) together with the predictions by the fits to the Bell-Evans model using Eq.(\ref{eq.dFT-CTF}b) (dashed line) and the MFA, Eq.(\ref{eq.JMA}a), assuming that $\log(c'/c)=0$ (solid line). 

In Figure \ref{fig5}a we test the full feedback-FT Eq.(\ref{eq.FT-CTF}a). For comparison we also show the non-feedback case. We emphasize the importance of properly weighing $\rho_\leftarrow \left( -W| \lambda \right)$ to build $\rho_\leftarrow(-W)$. An \textit{unweighted} reverse work distribution ($\int{\rho_\leftarrow\left(-W|\lambda\right)d\lambda}$, blue) does not fulfil the FT (inset, blue points), and the slope of the fitting line ($\sim$0.08) is far below 1. Figure \ref{fig5}b (top) shows $\Upsilon_\infty$ for different experimental conditions (black circles) and results obtained in simulations (gray squares) of a hairpin model (S2, Supp. Info.) compared to the theoretical values determined from Eq.(\ref{eq.FT-CTF}b) using the Bell-Evans model fits of Fig.\ref{fig4}c. Also, the MFA using Eq.\ref{eq.JMA}b is shown as a dashed line. In Figure \ref{fig5}b (bottom) we show the efficiencies $\eta_I$ and $\eta_M$ versus $r_U/r_F$ . As shown in Figure \ref{fig5}b dissipation reduction is larger for CTF as compared to DTF (for CTF $\Delta \langle W_d \rangle$ is not bounded by the Landauer limit $k_BT\log 2$). However, $\eta_M\sim -0.1$ is slightly negative as in DTF, showing that dissipation reduction does not necessarily improve free energy determination. In Figure \ref{fig5}c we plot the experimental free energy bias Eq.\eqref{eqBias} as a function of the number of pulling experiments $N$ at the conditions shown in panel b: as for DTF, we observe that the bias with feedback does not decrease relative to the non-feedback case (purple triangles). 

% /////////////////////////////////////////////////////////////
\subsection{Inefficient information-to-measurement conversion}
\label{sec:ineinfotomeas}
%
% ///////////////////////////////////

By reducing dissipation, feedback might be used to improve free energy prediction. Second law’s inequality Eq.\eqref{eqWdis} permits to reduce $\langle W_d\rangle$ with respect to the bound without feedback, $\langle W_d \rangle_0 \geq 0$. However, this is not true if the reduction in work is lower than the thermodynamic information: $\langle W_d \rangle + k_BT\Upsilon \geq \langle W_d \rangle_0$ and $\eta_M \leq 0$. Then, the Jarzynski bias for Eq.\eqref{eqBiasN1} increases with feedback undermining free energy determination \cite{Pala2011}. This is the case of the DTF and CTF experiments previously shown. 

Hairpin L4 exhibits low dissipation without feedback. Here we ask whether feedback efficiency increases upon increasing the irreversibility of the process ($\langle W_d \rangle_0$ larger). For this we have carried out numerical simulations of the phenomenological model for a new DNA hairpin (L8). L8 has the same stem of the previous hairpin (L4, Fig.\ref{fig1}a) but with an 8-bases loop. L8 shows larger dissipation compared to L4 when pulled under the same experimental CTF protocol (Fig.\ref{fig4}a). The results for the pulling curves, $\widetilde{\psi}$, $p_\leftarrow^U$ and the test of the full feedback-FT Eq.(\ref{eq.FT-CTF}a)  are shown in S3 in Supp. Info. for hairpins L4 and L8
% _______________________________ figure 6
\begin{figure}
    \centering
    \includegraphics{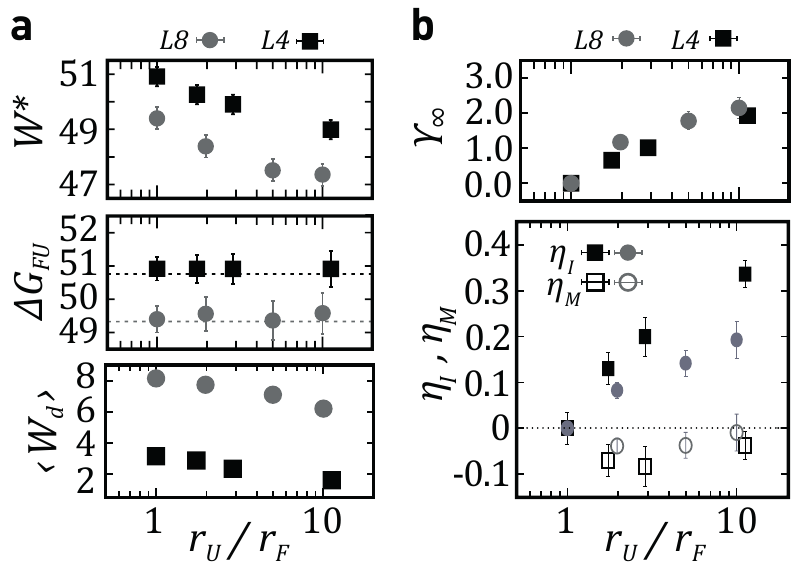}
    \caption{{\bf Free energy prediction with CTF.} (a) Values for $W^\ast$ (top), $\Delta G_{FU}$ (middle) and $\langle W_d\rangle$ (bottom) for hairpins L4 and L8. $\Delta G_{FU}$ values are obtained by adding $\Upsilon_\infty$ (panel b) and the crossing work values $W^\ast$, ${\Delta}G_{FU} = W^\ast + k_BT\Upsilon_\infty$, and are compared to the correct free energy values (dashed lines in ${\Delta}G_{FU}$). (b) Values for $\Upsilon_\infty$ Eq.(\ref{eq.FT-CTF}c), $\eta_I$ and $\eta_M$ Eqs.(\ref{eq.etaI},\ref{eq.etaM}) versus $r_U/r_F$. The values of $W^\ast$, $\langle W_d\rangle =\langle W \rangle - \Delta G_{FU}$ and $\Upsilon_\infty$ equally decrease and increase with $r_U/r_F$. All energy values are in $k_BT$ units.}
    \label{fig6}
\end{figure}
%_______________________________
% ___________________________________ Figure 7
\begin{figure*}
\centering\includegraphics{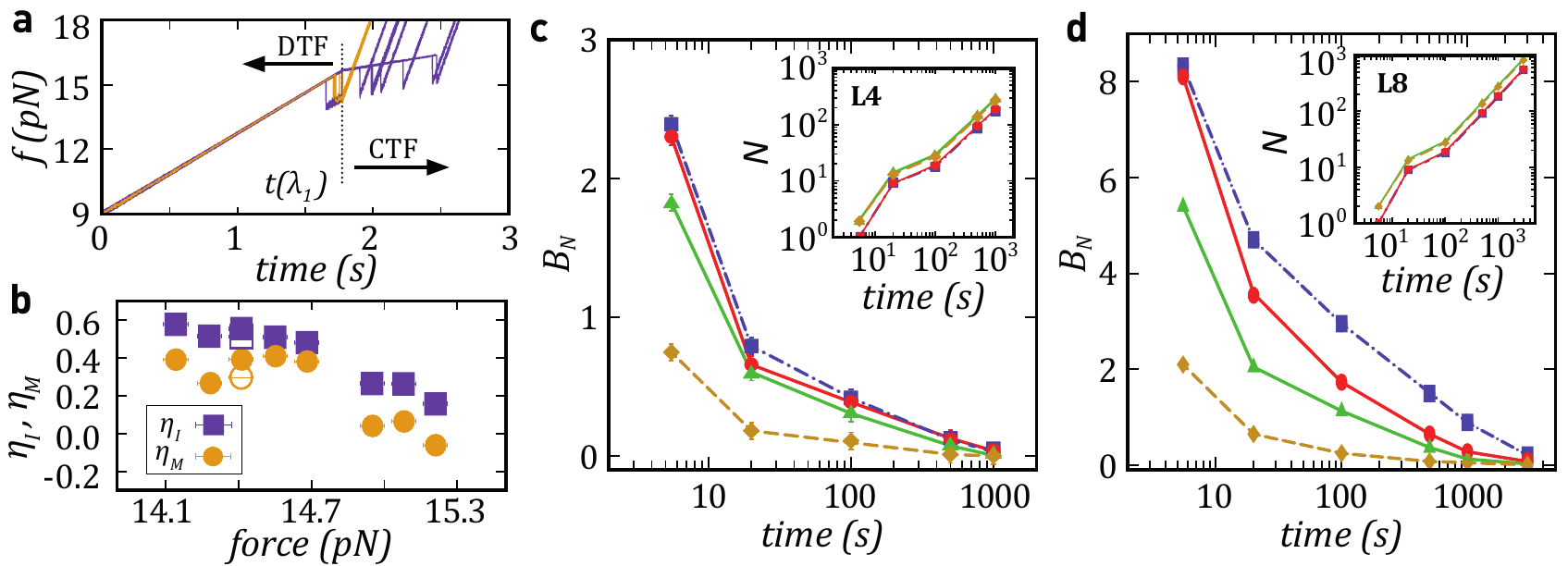}
    \caption{{\bf DTF+CTF strategy and bias versus total experimental time.} (a) Force versus time for the DTF + CTF strategy. The pulls start with DTF at $r_F = 4$pN/s and the measurement position is at $\lambda_1$. If the molecule is observed to be in U at $\lambda_1$, the loading rate is changed to $r_U = 17$pN/s (orange trajectories). If the molecule is observed to be in F at $\lambda_1$ the CTF is turned on with a loading rate $r'_F = 1$pN/s $<r_F$ (purple trajectories). At the first unfolding event, the loading rate is changed from $r'_F$ to $r_U$. (b) Efficiencies $\eta_I$ and $\eta_M$ for DTF+CTF. Solid symbols correspond to molecule L4, and empty symbols correspond to molecule L8. The values of $\Upsilon$ for this protocol were directly determined from the bias as we do not have an analytical expression for the thermodynamic information in this case. (c) and (d) Bias versus the experimental times measured from numerical simulations of molecule L4 (c) and L8 (d) with non-feedback (purple), DTF (red), CTF (green), and DTF+CTF (yellow) protocols using $r_F = 4$pN/s, $r_U = 17$pN/s, and $r'_F = 1$pN/s. Inset: Simulated trajectories (N) for a given total experimental time. The same plots versus N (for DTF+CTF) are shown in Fig. S3 in Supp. Info. }
    \label{fig7}
\end{figure*}
% _____________________________________________________________
Figure \ref{fig6} summarizes the main results obtained from the simulations for L4 and L8 under CTF (S3, Supp. Info.) by varying $r_U$ for a fixed $r_F$. We show: the values of $W^\ast = \Delta G_{FU} - k_BT \Upsilon_\infty$ (Fig.\ref{fig6}a, top) derived from the crossing point of forward and reverse work distributions from the full feedback-FT; the folding free energy of the hairpin predicted by the full feedback-FT Eq.(\ref{eq.FT-CTF}a) (Fig.\ref{fig6}a, middle) and; the average dissipated work (Fig.\ref{fig6}a, bottom). From these values we derive the thermodynamic information $\Upsilon_{\infty}$ and the efficiencies $\eta_I,\eta_M$ (Fig.\ref{fig6}b).  
It is worth noticing that both $W^\ast$ and $\langle W_d\rangle$ mildly decrease with $r_U/r_F$ (Fig.\ref{fig6}a, top and bottom). However the values of $W^\ast$ and ${\Upsilon}_\infty$ (Fig.\ref{fig6}b, top) compensate each other yielding fairly constant estimates for ${\Delta}G_{FU} = W^\ast + k_BT \Upsilon_\infty$ that are compatible with the values of ${\Delta}G_{FU}$ used in the simulations (Fig.\ref{fig6}b, middle). Despite the larger irreversibility of L8, dissipation reduction defined by $\Delta\langle W_d \rangle=\langle W_d \rangle_0-\langle W_d \rangle$ is similar for L4 and L8 (Fig.\ref{fig6}a, bottom). Moreover, also $\Upsilon_\infty$ (Fig.\ref{fig6}b, top) remains similar for both hairpins. This shows that increased irreversibility (quantified by $\langle W_d \rangle_0$) does not necessarily imply larger $\Delta\langle W_d \rangle$ and $\Upsilon_\infty$. In general, despite the fact that $\langle W_d \rangle$ decreases with positive feedback, the values of $W^\ast$ and $\Upsilon_\infty$ decrease and increase at the same rate, respectively. Therefore $\langle W_d \rangle + k_BT\Upsilon \cong \langle W_d \rangle_0 \geq 0$: feedback does not make the inequality imposed by the Second Law any weaker. Accordingly, $\eta_M$ remains $\cong 0$ for all $r_U/r_F$ values (Fig.\ref{fig6}b, bottom) indicating inefficient information-to-measurement conversion. Overall these results demonstrate that, although CTF reduces dissipation, this is compensated by an equal decrease of $W^\ast = \Delta G_{FU} - k_BT\Upsilon_\infty$, leading to $\eta_M \cong 0$ and unimprovement in free energy determination.

\subsection{Efficient information-to-measurement conversion: from protocols to strategies}
\label{sec:effiinfotomeas}
Here we ask under which conditions feedback does improve free energy determination increasing $\eta_M$. As previously shown for hairpin L8, the irreversibility of the non-feedback process barely changes  $\eta_M$. In CTF dissipation reduction is larger than for DTF, however this comes at the price of a larger $k_BT\Upsilon$, leading to $\eta_M\cong 0$.

Here we explore the possibility of modifying the feedback protocols in such a way that the dissipation reduction, $\Delta\langle W_d \rangle$, is maximized relative to $k_BT\Upsilon$. In non-feedback pulling experiments, dissipation reduction can be achieved by simply decreasing the loading rate  (i.e., making the process less irreversible). However, this comes at the price of an increase in the average time per pulling cycle and a decrease of the total number of pulls in a {\it day of experiments}, rendering free energy determination inefficient.  The interesting problem is to reduce dissipation with feedback, keeping the average time per pulling cycle equal or lower to the average time per pulling cycle without feedback.

In the DTF protocol, we increased the pulling rate only when the molecule was found to be in U at $\lambda_1$, while no action was taken if the molecule was in F.  As shown in Appendix \ref{sec:BellEvans} (Eq.\eqref{AWd_BE}) dissipation reduction is the product of the fraction of trajectories that are in U at $\lambda_1$, $p_\rightarrow^U (r_F)$, and the dissipated work reduction conditioned to the U-type trajectories. At high forces $p_\rightarrow^U (r_F)$ is large whereas dissipation reduction is low (Figures \ref{fig10}c,\ref{fig11}c). Conversely, $p_\rightarrow^U (r_F)$ is small at low forces where dissipation reduction is the largest. Maximal $\Delta\langle W_d \rangle$ is found close to the coexistence force where the terms $p_\rightarrow^U (r_F)$ and $\left[ \langle W_d \rangle_{U \rightarrow  F \rightarrow U}(r_U) - \langle W_d \rangle_{U \rightarrow  F \rightarrow U}(r_F) \right]$ balance. To further reduce dissipation one might consider applying feedback also to the large set of F-type trajectories at $\lambda_1$, e.g. by reducing the pulling rate after $\lambda_1$. 

To show that $\eta_M$ can be positive and large we have implemented a feedback strategy combining DTF and CTF. In this DTF+CTF strategy the molecule is initially pulled at $r_F$ with DTF until $\lambda_1$ where a observation is made. If the outcome is U then the pulling rate is switched to $r_U> r_F$ between $\lambda_1$ and $\lambda_{max}$. Instead, if the outcome is F the pulling rate is reduced to $r_F'< r_F$ and the CTF protocol turned on. In this case, at the first unfolding event after $\lambda_1$, the pulling rate is switched to $r_U> r_F>r_F'$ until $\lambda_{max}$. In the DTF+CTF protocol both U- and F-trajectories contribute to reduce the dissipated work. Moreover, the values of $r_F'$ can be chosen such that the average time per pulling trajectory is lower compared to the non-feedback case. In Figure \ref{fig7}a,b we show the results obtained for hairpin L4 in the DTF+CTF strategy where $r_F=4>r_F'=1$pN/s, $r_U=17$pN/s. In the coexistence force region ($f_1\simeq 14.5$pN) dissipated work is reduced by roughly 50$\%$ while $\Upsilon$ remains unchanged with respect to the standard CTF protocol, leading to $\eta_I\sim 0.6$, $\eta_M\sim 0.4$ (Fig.\ref{fig7}b). Similar results are obtained for L8 with same rates at one specific condition. We find that $B_1$ decreases by $\sim 1k_BT$ and $\sim 6k_BT$ for L4,L8 respectively (Section S4 in Supp. Info.). In addition, we compared the bias as a function of the total experimental time for the four studied protocols (non-feedback, DTF, CTF, and DTF+CTF) from numerical simulations using molecule L4 and L8 for the same pulling rates. In Figure \ref{fig7}c,d we show the time dependence of the bias, while in inset of Figures \ref{fig7}c,d we present the time dependence of the number $N$ of simulated trajectories. Although CTF generates the largest number of trajectories, the DTF+CTF strategy is the most efficient one.

\subsection{Efficiency plot}
\label{sec:effiplot}
\begin{figure}
\centering\includegraphics{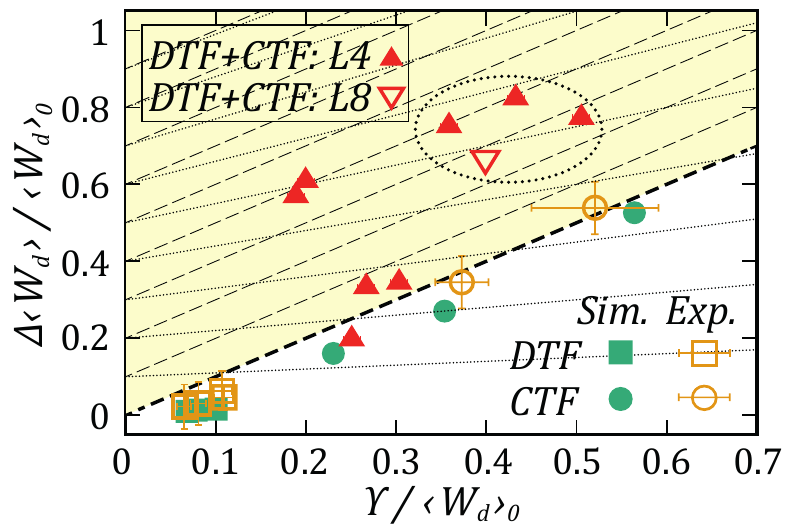}
    \caption{ {\bf Efficiency plot.}  Dissipation reduction, $\Delta \langle W_d\rangle$, versus thermodynamic information, $k_B T\Upsilon$, normalized by the non-feedback dissipation value, $\langle W_d\rangle_0$, for all explored cases in L4: CTF (empty circles, experimental data; full circles, simulated data), DTF (empty squares, experimental data; full square, simulated data) and the DTF+CTF strategy (red triangles). Dotted and dashed lines correspond to Eqs.(\ref{etaI_effiP},\ref{etaM_effiP}) and intersect the y-axis at $\eta_I,\eta_M$. Red empty triangles correspond to the $\Delta \langle W_d\rangle$ versus $k_B T\Upsilon$ for L8 molecule under DTF+CTF protocol.}
    \label{fig8}
\end{figure}

To show all results in perspective we introduce the efficiency plot (Figure \ref{fig8}). We plot the dissipation reduction $\Delta\langle W_d \rangle=\langle W_d \rangle_0-\langle W_d \rangle$ versus $k_BT\Upsilon$, both normalized by the non-feedback dissipation value $\langle W_d \rangle_0$ ($\sim 2k_BT$). Results are shown for hairpin L4 from experiments and simulations (yellow and green symbols), for DTF and CTF (squares and circles) and DTF+CTF (red triangles); and for L8 at the specific DTF+CTF condition shown in Fig.\ref{fig7}b. The black-dashed line $\Delta\langle W_d \rangle=k_BT\Upsilon$ separates two regions: $\eta_M>0$ (second law’s weakening, yellow region) and $\eta_M<0$ (second law’s strengthening, white region). Remarkably, despite of dissipation reduction $\Delta\langle W_d \rangle>0$, all results for DTF and CTF fall on the region $\eta_M\cong 0$ (squares and circles, dashed line), indicating that the second law is strengthened with feedback. Therefore a large dissipation reduction (rightmost green and yellow circles) does not necessarily imply free energy determination improvement (Sec.\ref{sec:ineinfotomeas}). To evaluate this we use the definitions of $\eta_I$ and $\eta_M$, Eqs.(\ref{eq.etaI},\ref{eq.etaM}), to express $\Delta\langle W_d \rangle/\langle W_d \rangle_0$ and $k_BT\Upsilon/\langle W_d \rangle_0$ as sole functions of $(\eta_I,\eta_M)$:
\begin{subequations}
\begin{align}
 \frac{\Delta \langle W_d\rangle}{\langle W_d\rangle_0} = \eta_I \frac{\eta_M - 1}{\eta_I-1} \\
   \frac{k_BT\Upsilon}{\langle W_d\rangle_0} = \frac{\eta_M - \eta_I}{\eta_I-1} \ ,
\end{align}
\end{subequations}
which for $\eta_M = 0$ gives $\Delta\langle W_d \rangle=k_BT\Upsilon$ (black dashed line in Fig.\ref{fig8}). Efficiencies $\eta_I,\eta_M$ separately define two linear relations between $\Delta\langle W_d \rangle/\langle W_d \rangle_0$ and $k_BT\Upsilon/\langle W_d \rangle_0$,
\begin{subequations}
\begin{align}
   \frac{\Delta \langle W_d\rangle}{\langle W_d\rangle_0} = \eta_I + \eta_I\frac{k_BT\Upsilon}{\langle W_d\rangle_0} \label{etaI_effiP}\\
   \frac{\Delta \langle W_d\rangle}{\langle W_d\rangle_0} = \eta_M +\frac{k_BT\Upsilon}{\langle W_d\rangle_0}\,. \label{etaM_effiP}
\end{align}
\end{subequations}
These are shown as dotted (Eq.\eqref{etaI_effiP}) and dashed (Eq.\eqref{etaM_effiP}) lines in Fig.\ref{fig8} of slopes equal to $\eta_I$ and 1, and intersections with the y-axis equal to $\eta_I,\eta_M$, respectively. For a given point in the efficiency plot we can read the values of $\eta_I,\eta_M$ by drawing lines of slopes $\eta_I$ and 1 to match the values $\eta_I,\eta_M$ in the y-axis. As we can see the DTF+CTF strategy yields the largest efficiencies for the largest $k_BT\Upsilon/\langle W_d \rangle_0$ values measured in CTF (circled region). The efficiency plot shows there is room for improved free energy prediction, opening the question of finding strategies that maximize $\eta_M$.

% ___________________________________ Figure 8
% _____________________________________________________________

% ____________________________________________________________________
%
\section{Conclusions}
We have investigated dissipation reduction and information-to-measurement conversion in DNA pulling experiments with feedback.  We have carried out irreversible pulling experiments on DNA hairpins that are mechanically folded and unfolded, finding conditions in which feedback does reduce dissipation. In the absence of feedback there is net  dissipated work and the Jarzynski free energy estimator is biased. We ask whether dissipation reduction can be used to improve free energy determination by weakening the second law inequality (i.e., by a reduction of the Jarzynski bias). We find that DTF and CTF protocols mildly reduce dissipation being highly inefficient for free energy determination. In contrast, a combination of the two protocols (denoted as a strategy) is much more efficient.

We have introduced cycle efficiencies $\eta_I$, $\eta_M$ for information-to-work (dissipation reduction, $\Delta \langle W_d\rangle>0$) and information-to-measurement (second-law inequality weakening, $\langle W_d \rangle+k_BT \Upsilon<\langle W_d \rangle_0$) in irreversible pulling experiments with discrete-time (DTF) and continuous-time feedback (CTF). These are particular cases of the first-time feedback (1$^{\rm st}$TF) protocol where the pulling rate $r_F$ switches to $r_U$ the first time the molecule unfolds along a predetermined sequence of $M$ measurement trap positions. A detailed and full feedback-FT has been derived for such a protocol that is expressed in terms of the free energy difference, ${\Delta}G_{FU}$, between the unfolded and folded states (Eqs.(\ref{eq.FT-1stTF}, \ref{eq.Ym-1stTF})), and in terms of two new quantities, namely the partial information $J_k$ and the full thermodynamic information $\Upsilon_M$. For $M=2$, 1$^{\rm st}$TF maps onto DTF, Eqs.(\ref{eq.FT-DTF}a, \ref{eq.FT-DTF}b), the case originally considered in Refs.\cite{Toy2010,Sag2010}. Applied to two-state molecules, DTF reduces dissipation by at most $k_BT\Upsilon_2 = k_BT\log{2}$ (Landauer limit). In the opposite case, $M\rightarrow\infty$, we obtain a novel work-FT for CTF Eqs.(\ref{eq.dFT-CTF}a, \ref{eq.FT-CTF}a) for the partial ($J(\lambda)$) and full thermodynamic information ($\Upsilon_\infty$), which is amenable to experimental test. Note that $\Upsilon_\infty$ is finite and unbounded, a consequence of the fact that the information-content of the stored sequences diverges. It is an open question the relation of $\Upsilon_M$ in Eq.\eqref{eq.Ym-1stTF} to other information-based related quantities \cite{Rol2014,Horo2014,Ford2016,Gav2017,Ali2019}. Interestingly, $\Upsilon_M$ is reminiscent of an equilibrium free energy potential, $G = -k_BT \log \sum_\sigma \exp \left( -G_\sigma / k_BT \right)$ with $G_\sigma$ the partial free energy of state $\sigma$. By defining $g_k = J_k + \log{\psi_k}$, Eq.\eqref{eq.Ym-1stTF} can be recast as a free energy, $\Upsilon_M = \log\sum_k \exp g_k$ indicating that thermodynamic information stands for a free energy difference.

We have carried out experiments for DTF and CTF on hairpin L4 for pulling rates in the same range $r_F\sim 4-5$pN/s, $r_U\sim 17-23$pN/s. The experiments have been complemented with numerical simulations of a phenomenological model for hairpins L4 and L8, and theoretical estimates of the Bell-Evans two-state model in the mean-field (MFA, Sec.\ref{sec:JMA}) and single-hopping (Appendix \ref{sec:BellEvans}) approximations. We find that CTF leads to higher $\Upsilon$ and $\eta_I$ compared to DTF (Figs.\ref{fig3}b,\ref{fig5}b). Indeed, CTF profits on early and rare unfolding events during the pulling protocol, making $\Upsilon$ and $\eta_I$ larger, a feature also observed in a recent experimental realization of the continuous equilibrium MD \cite{Rib2019,Rib2019_2}. In contrast, both DTF and CTF are inefficient regarding $\eta_M$: $\langle W_d \rangle$ decreases by roughly $k_BT\Upsilon$ leaving the second law inequality unweakened and the Jarzynski bias almost unchanged with feedback. In fact,  by strategically combining DTF and CTF we can make information-to-measurement conversion efficient (Figure \ref{fig7}a). The DTF+CTF strategy maximizes dissipated work reduction without increasing $k_BT\Upsilon$ leading to high $\eta_I,\eta_M$ values (Figure \ref{fig7}b). The results are summarized in the efficiency plot (Figure \ref{fig8}) which demonstrates that efficient information-to-measurement conversion is obtained by maximizing $\Delta \langle W_d\rangle/\langle W_d\rangle_0$ while minimizing $k_BT\Upsilon/\langle W_d\rangle_0$ (ideally becoming negative). Our results show that feedback strategies (defined as a set of multiple-correlated feedback protocols) enhance the information-to-measurement efficiency, opening the door to find optimal strategies for improved free energy determination.  

Information-to-measurement conversion might be interpreted as a two-steps process, with work reduction as an intermediate step of information-to-measurement conversion: information is first used to reduce work (information-to-work conversion, $\Delta \langle W_d\rangle>0$, efficiency $\eta_I$), followed by $\Delta G$ determination (work-to-measurement conversion, $k_BT\Upsilon$ small, efficiency $\eta'$). Therefore $\eta_M = \eta_I\eta' \le \eta_I,\eta' \le 1$. These two steps must be correlated to maximize the overall efficiency, requiring multiple-correlated feedback protocols.  

It would be interesting to search other non-equilibrium protocols or physical settings where $\eta_M(\lesssim\eta_I)$ is maximized. The vast majority of previous theoretical and experimental studies operate on systems that, in the absence of feedback, are in equilibrium. A handful of papers have studied dissipation reduction in non-equilibrium settings \cite{Abreu2012,tafoya2019}, but none of them have considered the information-to-measurement conversion. Dissipation reduction by feedback control has  also been studied in macroscopic systems, e.g. feedback cooling  \cite{poot2012}, electronic and logic circuits \cite{sheikhfaal2015} and climate change \cite{ozawa2003}. In general, feedback control corrects deviations from a reference state by monitoring the time evolution of a macroscopic observable, leading to higher dissipation. In contrast, dissipation reduction in small systems requires rectifying thermal fluctuations. It is in this context where feedback-FTs are applicable.

Future studies should also address information-to-measurement conversion in systems with measurement error \cite{Wach2016,Hof2018}, non-Markovian dynamics \cite{Muna2014,Deb2020}, biologically inspired \cite{Ito2015,Brit2019,Tang2020,Dab2019} and mutually interacting or autonomous systems \cite{Adm2018,Kos2014_2,Sag2012_2,Kos2015}. The latter might be tested designing single molecule constructs containing multiple DNA structures. These studies will enlarge our understanding of transfer energy and information flow in non-equilibrium systems.

\begin{acknowledgments}
We thank A. Alemany and J. Horowitz for their contribution in the initial stages of this work. R.K.S., J.J. and H.L. were supported by the Swedish Science Council (VR) project numbers 2015-04105, 2015-03824 Knut and Alice Wallenberg Foundation project 2016.0089. J.M.R.P acknowledges support from Spanish Research Council Grant FIS2017-83706-R. M.R. and F.R. acknowledge support from European Union’s Horizon 2020 Grant No. 687089, Spanish Research Council Grants FIS2016-80458-P, PID2019-111148GB-I00 and ICREA Academia Prizes 2013 and 2018.
\end{acknowledgments}

\iffalse
\section*{Author Contributions}
\textit{J. M. P., H. L., J. J. and F.R., planned the research, M. R.-P, M.R.-C and R.K.S. performed the experiments. M. R., and R.K.S. analysed the data and did numerical simulations. M.R.-P R.S., J.M.R.P and F.R. did the theoretical calculations. All authors contributed to the writing.}
\fi

% _____________________ APPENDIX _________________
\appendix 
\section{Derivation of the 1$^{\rm st}$TF-FT and the CTF limit}\label{sec:1stTF_der} 

In this section we present the derivation of the detailed and full work fluctuation theorem (work-FT) for the first-time feedback (1$^{\rm st}$TF) protocol. As a corollary, we derive the continuous-time feedback (CTF) limit. The derivation of the 1$^{\rm st}$ time-FT is an application of the extended version of Crooks-FT \cite{Cro1999} introduced in  \cite{Juni2009} to reconstruct free energy branches, and applied to derive free energies of kinetic states \cite{Ale2012}, and ligand binding \cite{Camu2017}.

In pulling experiments the force is ramped with a constant loading rate $r_F$. Measurements are made as a function of time or $\lambda$ (natural control parameter in our optical tweezers setup). In the 1$^{\rm st}$TF protocol measurements are taken at a pre-determined set of trap positions along the pulling curve, $\lbrace \lambda_k; 0\le k\le M\rbrace$, at given times $\lbrace t_k; 0\le k\le M\rbrace$ starting from an initial time, $t_0 = 0$, up to a final time $t_M$. Therefore there is a total number of $M-1$ observations made for each trajectory (the initial and final times are excluded) implying that $M\ge 2$.  
 
The force and $\lambda$ limits in pulling experiments are such that the molecule is always folded (F) at $\lambda_0$ and unfolded (U) at $\lambda_M$. This condition can be relaxed to include cases where the system starts either in equilibrium or in U at $\lambda_0$. However, for simplicity we will stick to the scenario applicable to the experiments where the molecule always starts in F at $\lambda_0$ and always ends in U at $\lambda_M$.  A measurement of the force is made at each $\lambda_k$ and the state of the molecule, F or U, is determined depending on whether it falls in the folded or unfolded branch ($f_F(\lambda),f_U(\lambda)$).  Therefore, each stochastic trajectory $\Gamma$ is defined by a sequence of F and U symbols, $\Gamma \equiv \lbrace F^0,.... F^{k^*-1}, U^{k^*}, ...,U^M\rbrace$ ($1\le k^* \le M$). The 1$^{\rm st}$TF protocol changes the loading rate from the initial value $r_F$ to a second value $r_U$ the first time $t_{k^*}$ an unfolding event is observed at $\lambda_{k^*}$. It is important to stress the notion of {\it first time} event. In the above trajectory $\Gamma$, the first part of the sequence of measurements until position $\lambda_k$, $\lbrace F^0,.... F^{k^*-1}\rbrace$, only contains $F$ symbols, whereas the second part between $\lambda_{k^*}$ and the limit $\lambda_M$, , $\lbrace U^{k^*},.... F^{k^M}\rbrace$, always starts in U at $\lambda_{k^*}$ and ends in U at $\lambda_M$ with either none or multiple (even) hopping transitions ($F\leftrightarrow U$) in between. 
For example, for $M=2$, there are two possible types of trajectories, $\lbrace F^0,F^1,U^2\rbrace$ and $\lbrace F^0,U^1,U^2\rbrace$, depending on the measurement outcome (F,U) at $\lambda_1$. $M=2$ corresponds to the discrete-time feedback case studied in the main text. Note that the number of different measurement sequences $\Gamma$ in the 1$^{\rm st}$TF protocol equals $2^{M-1}$. 

To derive the detailed work-FT, first we define the total forward work probability, $\rho_\rightarrow\left(W|k^*\right)$, conditioned to the first unfolding event taking place at $\lambda_{k^*}$ ($1\le k^* \le M$):
\begin{eqnarray}
\rho_\rightarrow\left(W|k^\ast\right) = \int \prod_{k=0}^{k^\ast-1} \left[ \rho_{\lambda_k \rightarrow \lambda_{k+1}}(W_{k+1})dW_{k+1} \right] \cdot \nonumber \\
\cdot dW' \rho_{\lambda_{k^\ast} \rightarrow \lambda_M}(W') \delta(W-\sum_{k=0}^{k^\ast-1}W_{k+1}-W') .
\label{rhoforward}
\end{eqnarray}
In Eq.\eqref{rhoforward} $\rho_{\lambda_k \rightarrow \lambda_{k+1}}(W_{k+1})$ is the forward work distribution for the section  $\lambda_k\rightarrow\lambda_{k+1}$ in the first part of the trajectory $0\le k \le k^*-1$ and $\rho_{\lambda_{k^*} \rightarrow \lambda_M}(W')$ is the forward work distribution for the second part of the trajectory, $\lambda_{k^*} \rightarrow\lambda_M$. Following the notation in the main text, $\lambda_0\equiv\lambda_{min}$ and $\lambda_M\equiv\lambda_{max}$.

Next, we apply the detailed work-FT \cite{Juni2009} to the work distributions measured along the two parts of trajectory $\Gamma$ (before and after the first unfolding event at $\lambda_{k^*}$) and define the corresponding reverse work distributions. Notice that the reverse process is the time reverse of the forward one, meaning that the unloading rate in the reverse process equals $r_U$ between $\lambda_M$ and $\lambda_{k^*}$ switching back to $r_F$ between $\lambda_{k^*}$ and $\lambda_0$. Notice there is no condition on the molecular state $\sigma_{k^*}$ along the reverse process $\lambda_{k^*}$. We have:
\begin{enumerate}
% Primer punt_______________________________________________________________
	\item Case $0 \leq k \leq k^*-1$ (First part of $\Gamma$):
\begin{eqnarray}
 \rho_{\lambda_k \rightarrow \lambda_{k+1}}(W_{k+1}) = 
 \rho_{\lambda_k \leftarrow \lambda_{k+1}} (-W_{k+1}) \cdot \nonumber \\
 \cdot \exp \left[ \frac{ W_{k+1}-\Delta G_{\sigma_k,\sigma_{k+1}} + k_BT \log \left( \frac{\phi^{\leftarrow}_{\sigma_k}}{\phi^{\rightarrow}_{\sigma_{k+1}}}   \right)}{k_BT}\right]
\label{rhopart1}
\end{eqnarray}
Here $\phi^{\rightarrow}_{\sigma_{k+1}}$ is the fraction of forward trajectories that end in state $\sigma_{k+1}$ at $\lambda_{k+1}$ conditioned to start in state $\sigma_k$ at $\lambda_k$. $\phi^{\leftarrow}_{\sigma_{k}}$ is the fraction of reverse trajectories that end in state $\sigma_k$ at $\lambda_{k}$ conditioned to start in state $\sigma_{k+1}$ at $\lambda_{k+1}$. All these quantities (forward and reverse) are measured at pulling rate $r_F$. $\Delta G_{\sigma_k,\sigma_{k+1}}=G_{\sigma_{k+1}}(\lambda_{k+1})-G_{\sigma_{k}}(\lambda_{k})$ is the partial free energy difference between states $\sigma_{k+1}$ at $\lambda_{k+1}$ and $\sigma_k$ at $\lambda_{k}$. The partial free energy $G_{\sigma}(\lambda)$ of any state ($\sigma=F,U$) at a given $\lambda$ equals $G_{\sigma}(\lambda)=-k_BT\log Z_{\sigma}(\lambda)$ where $Z_{\sigma}(\lambda)$ is the partition function restricted to the set of configurations of state $\sigma$ at the trap position $\lambda$. Notice that in this first part of $\Gamma$, $\sigma_k=F$ for $0\le k\le k^*-1$ and $\sigma_{k^*}=U$. 

% Segon punt_______________________________________________________________
\item Case $k^* \leq k \leq M$ (Second part of $\Gamma$):
\begin{eqnarray}
\rho_{\lambda_{k^*} \rightarrow \lambda_{M}}(W') = \rho_{\lambda_{k^*} \leftarrow \lambda_{M}}(-W') \cdot \nonumber \\
\cdot \exp\left[\frac{W'-\Delta G_{\sigma_{k^*},\sigma_M} + k_BT \log \left( \frac{\phi^{\leftarrow}_{\sigma_{k^*}}}{\phi^{\rightarrow}_{\sigma_{M}}} \right)}{k_BT}\right] 
\label{rhopart2}
\end{eqnarray}
Here $\phi^{\rightarrow}_{\sigma_{M}}$ is the fraction of forward trajectories that end in $\sigma_M$ at $\lambda_{M}$ conditioned to start in ${\sigma_{k^*}}$ at $\lambda_{k^*}$. $\phi^{\leftarrow}_{{\sigma_{k^*}}}$ is the fraction of reverse trajectories that end in ${\sigma_{k^*}}$ at $\lambda_{k^*}$ conditioned to start in $\sigma_M$ at $\lambda_{M}$. All these quantities (forward and reverse) are measured at pulling rate $r_U$.
Note that the molecule is always in U at $\lambda_{k^*}$ so ${\sigma_{k^*}}=U$. Moreover all pulls along the forward process end in the unfolded state at $\lambda_M$ so ${\sigma_M}=U$ and $\phi^{\rightarrow}_{\sigma_{M}}=1$ in Eq.\eqref{eq.Ym-1stTF}. Analogously, $\Delta G_{\sigma_{k^*},\sigma_M}=\Delta G_{U,U}=G_U(\lambda_M)-G_U(\lambda_{k^*})$ is the partial free energy difference between state U at $\lambda_M)$ and $\lambda_{k^*}$.
\end{enumerate}
In Eqs.(\ref{rhopart1}, \ref{rhopart2}) $k_B$ is the Boltzmann constant and $T$ is the temperature. In what follows, to lighten notation we keep $\sigma_k,{\sigma_{k^*}},{\sigma_M}$ as free variables, only at the end we replace them with $\sigma_k=F$ for $0\le k\le k^*-1$ and ${\sigma_{k^*}},{\sigma_M}=U$. Inserting Eqs.(\ref{rhopart1}, \ref{rhopart2}) into Eq.\eqref{rhoforward} leads
\begin{eqnarray}\label{rhoforward2}
\rho_{\rightarrow}(W|k^*) = \int \!\left[ \prod_{k=0}^{k^*-1}dW_{k+1} \right] \cdot \nonumber \\
\cdot dW'\delta (W-\sum_{k=0}^{k^*-1}W_{k+1}-W') \cdot A \cdot B
\end{eqnarray} 
where $\beta = 1/k_BT$ and,
\begin{eqnarray}
 B &=& \prod\limits_{k=0}^{k^*-1} \rho_{\lambda_k \leftarrow \lambda_{k+1}}(-W_{k+1}) \rho_{\lambda_{k^*} \leftarrow \lambda_M}(-W')\label{B}\\
 A &=&  \prod_{k=0}^{k^*-1} \ e^{\beta \left[ W_{k+1}-\Delta G_{\sigma_k,\sigma_{k+1}} + k_BT \log \left( \frac{\phi^{\leftarrow}_{\sigma_k}}{\phi^{\rightarrow}_{\sigma_{k+1}}} \right)  \right]} \cdot  \nonumber \\ 
 & & \cdot e^{\beta \left[ W'-\Delta G_{\sigma_{k^*},\sigma_M} + k_BT \log \left( \frac{\phi^{\leftarrow}_{\sigma_{k^*}}}{\phi^{\rightarrow}_{\sigma_{M}}} \right)  \right]} = \nonumber\\
 &=& A' \cdot A'' \cdot A'''\label{A}
\end{eqnarray}
with
\begin{eqnarray}
 A' &=& \prod_{k=0}^{k^*-1} e^{\beta W_{k+1} } \cdot e^{\beta W'} = e^{\beta \left( \sum_{k=0}^{k^
 *-1}W_{k+1}+W' \right)}  = \nonumber \\
 &=& e^{\beta W}\label{a1}\\
 A'' &=& \prod_{k=0}^{k^*-1} e^{-\beta \Delta G_{\sigma_k,\sigma_{k+1}} } \cdot e^{-\beta \Delta G_{\sigma_{k^*},\sigma_{M}}} = \nonumber \\
 &=& e^{-\beta \sum \limits_{k=0}^{k^*-1} \Delta G_{\sigma_k,\sigma_{k+1} }- \beta \Delta G_{\sigma_{k^*},\sigma_M}} = e^{-\beta \Delta G_{F,U}}\label{a2}\\
 A''' 
&=&\prod_{k=0}^{k^*-1} e^{\beta k_BT \log \left( \frac{\phi^{\leftarrow}_{\sigma_k}}{\phi^{\rightarrow}_{\sigma_{k+1}}}  \right)} \cdot e^{\beta k_BT \log \left( \frac{\phi^{\leftarrow}_{\sigma_{k^*}}}{\phi^{\rightarrow}_{\sigma_{M}}}  \right)} = \nonumber \\
&=& \prod_{k=0}^{k^*-1} \frac{\phi^{\leftarrow}_{\sigma_k}}{\phi^{\rightarrow}_{\sigma_{k+1}}}\cdot \frac{\phi^{\leftarrow}_{\sigma_{k^*}}}{\phi^{\rightarrow}_{\sigma_{M}}}
 = \frac{\prod\limits_{k=0}^{k^*-1}\phi^{\leftarrow}_{\sigma_k}\cdot\phi^{\leftarrow}_{\sigma_{k^*}} } {\prod\limits_{k=0}^{k^*-1}\phi^{\rightarrow}_{\sigma_{k+1}}\cdot\phi^{\rightarrow}_{\sigma_{M}}} = \nonumber\\
&=& \frac{\prod\limits_{k=0}^{k^*-2}\phi^{F,F}_{\lambda_k \leftarrow\lambda_{k+1}}}{\prod\limits_{k=0}^{k^*-2} \phi^{F,F}_{\lambda_k \rightarrow\lambda_{k+1}}} \cdot \frac{\phi^{U,U,r_F}_{\lambda_{k^*\leftarrow \lambda_M}}}{\phi^{U,U,r_F}_{\lambda_{k^*\leftarrow \lambda_M}}} \cdot \frac{\phi^{F,U}_{\lambda_{k^*-1}\leftarrow \lambda_{k^*} }\phi^{U,U}_{\lambda_{k^*}\leftarrow\lambda_M}}{\phi^{F,U}_{\lambda_{k^*-1}\rightarrow \lambda_{k^*}} \phi^{U,U}_{\lambda_{k^*}\rightarrow\lambda_M}} 
= \nonumber \\
&=& \frac{\widetilde{\psi}_{k^*}}{\psi_{k^*}}\cdot\frac{p_{\leftarrow,k^*}^U(r_U)}{p_{\leftarrow,k^*}^U(r_F)}\label{a3}
\end{eqnarray}
where in Eqs.(\ref{a2}, \ref{a3}) we used $\sigma_0=F$, $\sigma_M=U$ and where we have introduced in the last line of Eq.\eqref{a3} a multiplicative factor equal to 1 ($\phi^{U,U,r_F}_{\lambda_{k^*\leftarrow \lambda_M}}/\phi^{U,U,r_F}_{\lambda_{k^*\leftarrow \lambda_M}} $). Moreover, in the last line of Eq.\eqref{a3} we adopted a specific notation for the conditional probabilities or fractions $\phi^{\leftarrow}_{\sigma_{k}}$, $\phi^{\rightarrow}_{\sigma_{k+1}}$, $\phi^{\leftarrow}_{\sigma_{k^*}}$ and $\phi^{\rightarrow}_{\sigma_{M}}$ previously introduced in Eqs.(\ref{rhopart1},\ref{rhopart2}), Table \ref{tab.2}. The conditional probabilities are:
\begin{enumerate}
    \item  $\phi^{F,F}_{\lambda_k \leftarrow\lambda_{k+1}}$ is the fraction of reverse trajectories where $\sigma_k=F$ at $\lambda_k$ conditioned to $\sigma_{k+1}=F$ at $\lambda_{k+1}$. This fraction is measured  with the unloading rate $r_F$.
    
    \item $\phi^{F,F}_{\lambda_k \rightarrow\lambda_{k+1}}$ is the fraction of forward trajectories where $\sigma_{k+1}=F$ at $\lambda_{k+1}$ conditioned to $\sigma_k=F$ at $\lambda_k$. This fraction is measured with the loading rate $r_F$.
    
    \item $\phi^{U,U,r_F}_{\lambda_{k^*}\leftarrow {\lambda_M}}$ is the fraction of reverse trajectories where $\sigma_{k^*}=U$ at $\lambda_{k^*}$ starting at $\sigma_M\equiv U$ at $\lambda_M$. As explicitly indicated in the notation, this fraction is measured at the unloading rate  $r_F$.
    
    \item  $\phi^{F,U}_{\lambda_{k^*-1}\leftarrow \lambda_{k^*}}$ is the fraction of reverse trajectories where $\sigma_{k^*-1}=F$ at $\lambda_k$ conditioned to $\sigma_{k^*}=U$ at $\lambda_{k^*}$. This fraction is measured  with the unloading rate $r_F$
    
    \item $\phi^{F,U}_{\lambda_{k^*-1}\rightarrow \lambda_{k^*}}$ is the fraction of forward trajectories where $\sigma_{k^*}=U$ at $\lambda_{k^*}$ conditioned to $\sigma_{k^*-1}=F$ at $\lambda_{k^*-1}$. This fraction is measured  with the unloading rate $r_F$;
    
    \item  $\phi^{U,U}_{\lambda_{k^*}\leftarrow \lambda_{M}}$ is the fraction of reverse trajectories where $\sigma_{k^*}=U$ at $\lambda_{k^*}$ starting at $\sigma_M\equiv U$ at $\lambda_M$. This fraction is measured with the unloading rate $r_U$.
    
    \item $\phi^{U,U}_{\lambda_{k^*}\rightarrow \lambda_{M}}$ is the fraction of forward trajectories where $\sigma_{M}=U$ at $\lambda_{M}$ conditioned to $\sigma_{k^*}=U$ at $\lambda_{k^*}$. This fraction is measured with the loading rate $r_U$. Note that, because all trajectories end in U at $\lambda_M$ the fraction $\phi^{U,U}_{\lambda_{k^*}\rightarrow \lambda_{M}}$  equals 1. 
\end{enumerate}
Note that in items 4 and 5 we have introduced the quantities $\phi^{F,U}_{\lambda_{k^*-1}\leftarrow \lambda_{k^*}}$ and $\phi^{F,U}_{\lambda_{k^*-1}\rightarrow \lambda_{k^*}}$, both measured with the loading rate $r_F$. 

{\renewcommand{\arraystretch}{2}
\begin{table*}
    \centering
    \begin{tabular}{c|l|l|c}element&
 notation in Eq.\eqref{a3} & notation in  Eq.(\ref{rhopart1},\ref{rhopart2}) & quantity measured at \\ \hline 
 \hline
1& $\phi^{F,F}_{\lambda_{k}\leftarrow \lambda_{k+1}}$  & $\phi^{\leftarrow}_{\sigma_{k}}$ with $\sigma_k=\sigma_{k+1}=F$ & $r_F$\\ \hline
 2&$\phi^{F,F}_{\lambda_{k}\rightarrow \lambda_{k+1}}$  & $\phi^{\rightarrow}_{\sigma_{k+1}}$ with $\sigma_k=\sigma_{k+1}=F$ & $r_F$\\ \hline
 3&$\phi^{U,U,r_F}_{\lambda_{k*}\leftarrow \lambda_{M}}$  & $\phi^{\leftarrow}_{\sigma_{k^*}}$ with $\sigma_k*=\sigma_{M}=U$ & $r_F$ \\ \hline
 4&$\phi^{F,U}_{\lambda_{k^*-1}\leftarrow \lambda_{k*}}$  &  $\phi^{\leftarrow}_{\sigma_{k^*-1}}$ with $\sigma_{k^*-1}=F;\sigma_{k^*}=U$& $r_F$\\ \hline
 5&$\phi^{F,U}_{\lambda_{k^*-1}\rightarrow \lambda_{k*}}$  &  $\phi^{\rightarrow}_{\sigma_{k^*}}$ with $\sigma_{k^*-1}=F;\sigma_{k^*}=U$& $r_F$\\ \hline
 6&$\phi^{U,U}_{\lambda_{k^*}\leftarrow \lambda_{M}}$  & $\phi^{\leftarrow}_{\sigma_{k*}}$ with $\sigma_k*=\sigma_{M}=U$& $r_U$ \\ \hline
 7&$\phi^{U,U}_{\lambda_{k*}\rightarrow \lambda_{M}}$(=1)  & $\phi^{\rightarrow}_{\sigma_{M}}$ with $\sigma_k^*=\sigma_{M}=U$ & $r_U$\\ \hline
        \end{tabular}
    \caption{\textbf{Notations.}}
    \label{tab.2}
\end{table*}}
To demonstrate the last equality in the last line of Eq.\eqref{a3} we group into a single product all fractions regarding reverse transitions at the unloading rate $r_F$ in the numerator ($\phi^{F,F}_{\lambda_k \leftarrow\lambda_{k+1}}$, $\phi^{F,U}_{\lambda_{k^*-1}\leftarrow \lambda_{k^*}}$, $\phi^{U,U,r_F}_{\lambda_{k^*}\leftarrow {\lambda_M}}$), and all fractions of forward transitions at the unloading rate $r_F$ in the denominator ($\phi^{F,F}_{\lambda_k\rightarrow\lambda_{k+1}}$, $\phi^{F,U}_{\lambda_{k^*-1}\rightarrow \lambda_{k^*}}$). We define
\begin{eqnarray}
& &\prod\limits_{k=0}^{k^*-2}\phi^{F,F}_{\lambda_k \rightarrow\lambda_{k+1}} \cdot\phi^{F,U}_{\lambda_{k^*-1}\rightarrow \lambda_{k^*}} \phi^{U,U}_{\lambda_{k^*\rightarrow \lambda_M}}=\psi_{k^*} \nonumber \\
& & \prod\limits_{k=0}^{k^*-2}\phi^{F,F}_{\lambda_k \leftarrow\lambda_{k+1}}\cdot \phi^{F,U}_{\lambda_{k^*-1}\leftarrow \lambda_{k^*}} \phi^{U,U,r_F}_{\lambda_{k^*\leftarrow \lambda_M}}=\widetilde{\psi}_{k^*}.
\label{psis}
\end{eqnarray}
$\psi_{k^*}(1\le k^* \le M)$  is the fraction of forward trajectories that start in F at $\lambda_0$ and are observed to be in U for the {\bf first} time at $\lambda_{k^*}$ with loading rate $r_F$. $\widetilde{\psi}_{k^*}(1\le k^* \le M)$ is the fraction of reverse trajectories that start in U at $\lambda_M$ and are observed to be in U for the {\bf last} time at $\lambda_{k^*}$ with unloading rate $r_F$. Note that in \eqref{psis} we have introduced the innocuous term for $\psi_{k^*}$, $\phi^{U,U}_{\lambda_{k^*\rightarrow \lambda_M}}=1$, in such a way that $\psi_{k^*}$ is the probability of full sequences $\Gamma$ (from $k=0$ to $M$) fulfilling the first-time condition. For $k^*=1$ the products $\prod\limits_{k=0}^{k^*-2}$ in Eq.\eqref{psis} are equal to 1. We stress two facts: 1) both $\psi_{k^*},\widetilde{\psi}_{k^*}$ are fractions measured at the single pulling rate $r_F$ without feedback and; 2) the notion of {\it first and last time} is bound to trajectories $\Gamma$  defined as sequences of observations at the pre-determination measurement positions $\lambda_k$ as they are defined in the 1$^{\rm st}$TF protocol, irrespective of what is the state of the molecule at other intermediate (unobserved) positions. 

Finally, in Eq.\eqref{a3}, $\phi^{U,U} _{\lambda_{k^*} \leftarrow \lambda_M} = p_{\leftarrow,k^*}^U(r_U)$ is the fraction of reverse trajectories that start in U at $\lambda_M$ and are observed to be in U at $\lambda_{k^*}$ at the unloading rate $r_U$. According to this definition we also have $\phi^{U,U,r_F}_{\lambda_{k^*}\leftarrow {\lambda_M}}=p_{\leftarrow,k^*}^U(r_F)$, a term which also appears in the denominator of the last fraction in Eq.\eqref{a3}.

Inserting Eqs.(\ref{a1},\ref{a2},\ref{a3}) in Eq.\eqref{A} and then in Eq.\eqref{rhoforward2} we notice that $A$ can be taken out of the integral Eq.\eqref{rhoforward2}. The remaining integral in Eq.\eqref{rhoforward2} contains only the term $B$ from Eq.\eqref{B}, which yields the reverse work distribution $\rho_{\leftarrow}(-W|k^*)$. We stress that the reverse work distribution is conditioned to forward process, through the first unfolding event observed at $\lambda_{k^*}$ along that process. Putting everything together we get the detailed work-FT for the 1$^{\rm st}$TF protocol,
 \begin{eqnarray}
  & &\boxed{ \frac{\rho_{\rightarrow}(W|k^*)}{\rho_{\leftarrow}(-W|k^*)} = \exp \left[ \beta\left( W - \Delta G_{F,U} + k_BT J_{k^*} \right) \right]} \nonumber \\
  & &\boxed{J_{k^*} = \log \left( \frac{\widetilde{\psi}_{k^*}} {\psi_{k^*}}\cdot\frac{p_{\leftarrow,k^*}^U(r_U)}{p_{\leftarrow,k^*}^U(r_F)}\right) (1\le k^*\le M)}
  \label{detailedWFT}
 \end{eqnarray}
Equation \eqref{detailedWFT} is the main theoretical result in this paper. $J_{k^*}$ is denoted as partial thermodynamic information and depends on four basic quantities ($\psi_{k^*},\widetilde{\psi}_{k^*}$, $p_{\leftarrow,k^*}^U(r_F)$, $p_{\leftarrow,k^*}^U(r_U)$). These quantities can be measured in protocols without feedback at the pulling rate $r_F$ along the forward process ($\psi_{k^*}$) and the reverse process ($\widetilde{\psi}_{k^*}$, $p_{\leftarrow,k^*}^U(r_F)$), and at the pulling rate $r_U$ along the reverse process ($p_{\leftarrow,k^*}^U(r_U)$).

From Eq.\eqref{detailedWFT} we derive the full work-FT for the 1$^{\rm st}$TF protocol.  The full work-distribution in the forward process is given by: 
\begin{eqnarray}
& &\rho_\rightarrow (W)=\sum_{k=1}^M \rho_\rightarrow(W|k)\cdot \psi_k = \nonumber\\
&=&\sum_{k=1}^M \rho_\leftarrow(-W|k)e^{\beta(W-\Delta G_{FU}+k_BT J_k)}\psi_k = \nonumber\\
&=&e^{\beta(W-\Delta G_{FU})}\cdot\sum_{k=1}^M \rho_\leftarrow(-W|k)e^{J_k} \cdot \psi_k = \nonumber\\
&=&e^{\beta(W-\Delta G_{FU})}\cdot\frac{\sum_{k=1}^M \rho_\leftarrow(-W|k)e^{J_k}\psi_k}{\sum_{k=1}^M e^{J_k}\cdot\psi_k}\cdot \nonumber \\
& &\cdot \sum_{k=1}^M e^{J_{k}} \cdot \psi_k
\end{eqnarray}
where, in the last line we have multiplied and divided by the term $\sum_{k=1}^M e^{J_{k}}\cdot \psi_k$. This allows us to define the reverse full-work distribution for the 1$^{\rm st}$TF protocol,
\begin{eqnarray}
\rho_\leftarrow(-W)&=&\frac{\sum_{k=1}^M \rho_\leftarrow(-W|k)e^{J_k+\log \psi_k}}{\sum_{k=1}^M e^{J_k+\log\psi_k}}\,\,\,.
\label{rhoreverse}
\end{eqnarray}
Notice that $\rho_\leftarrow(W)$ is properly normalized.
Finally we get:
\begin{equation}
\boxed{\frac{\rho_\rightarrow(W)}{\rho_\leftarrow (-W)}=e^{\beta(W-\Delta G_{FU}+k_BT\Upsilon_M)}}
\label{fullWFT}
\end{equation}
which is Eq.\eqref{eq.FT-1stTF} in the main text. The term $\Upsilon_M$ is the thermodynamic information and equals
\begin{equation*}
\Upsilon_M = \log\left[\sum_{k=1}^{M}{\psi_k e^{J_k}}\right]= \log\left[\sum_{k=1}^M\psi_k \frac{p_{\leftarrow,k}^U(r_U)}{p_{\leftarrow,k}^U(r_F)}\cdot\frac{\widetilde{\psi}_{k}}{\psi_k}\right]
\end{equation*}
\begin{equation}
\boxed{\Upsilon_M = \log\left(\sum_{k=1}^M \frac{p_{\leftarrow,k}^U(r_U)}{p_{\leftarrow,k}^U(r_F)}\widetilde{\psi}_{k}\right)}
\label{fullinfo}
\end{equation}
which gives Eq.\eqref{eq.Ym-1stTF} in the main text.
\\
\\
To conclude this section, we consider the continuous-time feedback (CTF) case corresponding to the limit $M\to\infty$ and determine the partial and full thermodynamic information, $J(\lambda)$ and $\Upsilon_\infty$, in such case. 

In this limit Eqs.(\ref{rhoforward}-\ref{fullinfo}) hold but with the continuous variable $\lambda$ replacing the discrete variable $k$. The partial thermodynamic information $J_k$ becomes the continuous function $J(\lambda)$ defined as:
\begin{equation}
\boxed{J(\lambda) = \log\left( \frac{p_{\leftarrow}^U(\lambda,r_U)}{p_{\leftarrow}^U(\lambda,r_F)}\frac{\widetilde{\psi}(\lambda)}{\psi(\lambda)}\right)} \ .
\label{partialinfoCFT}
\end{equation}
with equivalent definitions for the continuous fractions ($\psi(\lambda),\widetilde{\psi}(\lambda),p_{\leftarrow}^U(\lambda,r_U),p_{\leftarrow}^U(\lambda,r_F)$). The full thermodynamic information $\Upsilon_\infty$ is determined by taking the continuous limit, $\lambda_{k+1} = \lambda_k + \Delta \lambda$, where $\Delta \lambda \rightarrow 0$, and writing the sum in Eq.\eqref{fullinfo} as an integral:
\begin{equation}
\boxed{\Upsilon_\infty = \log\left(\int_{\lambda_{min}}^{\lambda_{max}} \frac{p_{\leftarrow}^U(\lambda,r_U)}{p_{\leftarrow}^U(\lambda,r_F)}\widetilde{\psi}(\lambda)d\lambda\right)}.
\label{fullinfoCFT}
\end{equation}
Equations (\ref{detailedWFT},\ref{fullWFT},\ref{partialinfoCFT},\ref{fullinfoCFT}) for the continuous-time limit of CTF yield Eqs.(\ref{eq.FT-CTF}a,\ref{eq.FT-CTF}b) in the main text.  

%/////////////////////////////////////////////////////////////////////

\section{FT for discrete-time feedback (DTF,$M = 2$)} 
\label{sec:DTF_der}
In this section we derive the FT for discrete-time feedback (DTF), i.e., Equations (\ref{eq.dFT-DTF},\ref{eq.FT-DTF}a) in the main text. The derivation is done with two methods, either as the 1$^{\rm st}$TF-FT for $M=2$ or by directly applying the extended-FT \cite{Juni2009} by classifying trajectories according to the measurement outcome at the intermediate position  $\lambda_1$.

We start by considering the 1$^{\rm st}$TF-FT for $M=2$. In this case state measurement sequences are of the type $\Gamma=\lbrace \sigma_0=F,\sigma_1,\sigma_2=U\rbrace$ corresponding to the three different measurements trap positions ($\lambda_k$): $\lambda_0$, $\lambda_1$ and $\lambda_2$. The relevant quantities in Eqs.(\ref{detailedWFT}, \ref{fullWFT}, \ref{fullinfo}) are $\psi_k$, $\widetilde{\psi}_k$, $p^{U}_{\leftarrow,k}(r_F)$, $p^{U}_{\leftarrow,k}(r_U)$ for $k=1,2$:
\begin{enumerate}
% segon element_____________________________ 
 \item Case $k=1$ ($\lambda_1$): In this case $p_{\leftarrow,1}^{U}(r_F) \neq 0$ and $p_{\leftarrow,1}^{U}(r_U) \neq 0$. Moreover, the fact that the molecule always starts in F (U) and ends in U(F) during the forward (reverse) process implies that the probability to observe the first (last) unfolding (refolding) event at $\lambda_1$ equals the probability that the molecule is in U at $\lambda_1$ during the forward (reverse) process. This holds for all pulling rate values: $\psi_1 = p_{\rightarrow,1}^{U}(r_F)$, $\widetilde{\psi}_1 = p_{\leftarrow,1}^{U}(r_F)$.

% tercer element_____________________________ 
 \item Case $k=2$ ($\lambda_2 \equiv \lambda_{max}$). By definition the probability to be in U at $\lambda_2$ equals 1 because all forward (reverse) trajectories end (start) in U, i.e., $p_{\leftarrow,2}^{U}(r_F) = p_{\leftarrow,2}^{U}(r_U) = 1$. Moreover,the fact that the molecule always starts in F (U) and ends in U (F) during the forward (reverse) process implies that the probability to observe the first (last) unfolding (refolding) event at $\lambda_2$ equals the probability that the molecule is in F at $\lambda_1$ during the forward (reverse) process. This holds for all pulling rate values:  $\psi_2 = p_{\rightarrow,1}^{F}(r_F)$, $\widetilde{\psi}_2 = p_{\leftarrow,1}^{F}(r_F)$.
\end{enumerate}
The different values of $p_{\leftarrow,k}^U$, $\psi_k$, $\widetilde{\psi}_k$ $(k=1,2)$ are presented in Table\,\ref{tab.3}.
% _________ tabla DTF
{
\renewcommand{\arraystretch}{2}
\begin{table}[h]
    \centering
    \begin{tabular}{c|c}
 $\lambda_1$ & $\lambda_2$ \\ \hline \hline
 $p_{\leftarrow,1}^U(r_F) \neq 0$  & $p_{\leftarrow,2}^U(r_F) = 1$  \\ \hline
 $p_{\leftarrow,1}^U(r_U)  \neq 0$  & $p_{\leftarrow,2}^U(r_U)  = 1$ \\ \hline
 $\psi_1 = p_{\rightarrow,1}^{U}(r_F) $  & $\psi_2 = p_{\rightarrow,1}^{F}(r_F)$\\ \hline
 $\widetilde{\psi}_1 = p_{\leftarrow,1}^U(r_F) $  & $\widetilde{\psi}_2 = p_{\leftarrow,1}^{F}(r_F)$\\ \hline
        \end{tabular}
    \caption{\textbf{Fractions for DTF.}}
    \label{tab.3}
\end{table}}
\\
From the results presented in Table\,\ref{tab.3} we calculate the partial and full thermodynamic information, $J_1\equiv I_U,J_2\equiv I_F$ (Eq.\eqref{detailedWFT}) and $\Upsilon_2$ (Eq.\eqref{fullinfo}):
\begin{eqnarray}
 \boxed{ I_U =J_1=\log\left( \frac{p_{\leftarrow,1}^{U}(r_U)}{p_{\rightarrow,1}^{U}(r_F)}\right)} \nonumber \\
 \boxed{I_F =J_2= \log\left( \frac{p_{\leftarrow,1}^{F}(r_F)}{p_{\rightarrow,1}^{F}(r_F)}\right)} \nonumber \\
 \boxed{ \Upsilon_2 = \log \left(p_{\leftarrow,1}^{U}(r_U) + p_{\leftarrow,1}^{F}(r_F)\right) }
\end{eqnarray}
Notice that $\rho_{\rightarrow}(W|k^*)$ in Eq.\eqref{detailedWFT} with $k^*=1,2$ corresponds to $\rho_{\rightarrow}(W|\sigma)$ in Eq.\eqref{eq.dFT-DTF} in the main text with $\sigma=U,F$, respectively. Therefore, Eq.\eqref{detailedWFT} for $k=1,2$ gives Eq.\eqref{eq.dFT-DTF} in the main text for $\sigma=U,F$, respectively. This completes the proof that the 1$^{\rm st}$TF-FT for $M=2$ equals DTF, Eqs.(\ref{eq.dFT-DTF}, \ref{eq.FT-DTF}a) in the main text.
%_________________________________________
\subsection{Alternative derivation of DTF}
Alternatively we can derive the detailed and full work-FT in DTF by classifying the trajectories in two classes depending on the observation made at $\lambda_1$ : 1) the system is in $F$ at $\lambda_1$ or, 2) the system is in U at $\lambda_1$. We use the extended-FT \cite{Juni2009} to calculate the detailed work-FT for each class of trajectories, first between $\lambda_0=\lambda_{min}$ and $\lambda_1$, next between $\lambda_1$ and $\lambda_2=\lambda_{max}$. These results are then combined to extract the detailed work-FT, for each class of trajectories, for the full pulling cycle between  $\lambda_0=\lambda_{min}$ and $\lambda_2=\lambda_{max}$.

The detailed work-FT, in the range $\lambda_0 \rightarrow \lambda_1$,  is given by \cite{Juni2009}: 
\begin{subequations}
    \begin{align}
       \frac{p^F_{\rightarrow}(\lambda_1)} {p^F_\leftarrow(\lambda_1)} \frac{\rho_{{\lambda_0},{\lambda_1}}(W|F)}{\rho_{{\lambda_0},{\lambda_{1}}}(-W|F)} = \exp\left(\beta (W - \Delta G_{F,F}^{\lambda_0,\lambda_1})\right)  \label{pFF}\\ \nonumber\\
           \frac{p^U_{\rightarrow}(\lambda_1)} {p^U_\leftarrow(\lambda_1)} \frac{\rho_{{\lambda_0},{\lambda_1}}(W|U)}{\rho_{{\lambda_0},{\lambda_{1}}}(-W|U)} = \exp\left(\beta(W - \Delta G_{F,U}^{\lambda_0,\lambda_1})\right)\label{pFU}
    \end{align}
\end{subequations} 
where $\Delta G_{F,F(U)}^{\lambda_0,\lambda_1}$ is the free energy difference between state F(U) at $\lambda_1$ and state F at $\lambda_0$. $\rho_{{\lambda_0},{\lambda_1}}(W|F(U))$ is the work distribution for the class of forward trajectories that start in F at $\lambda_0$ and end in F(U) at $\lambda_1$.  $\rho_{{\lambda_0},{\lambda_1}}(-W|F(U))$ is the corresponding reverse work distribution. $p^{F(U)}_{\rightarrow}(\lambda_1)$ are the probabilities in the forward process to be in F(U) at $\lambda_1$ conditioned to start in F at $\lambda_0$. $p^{F(U)}_{\leftarrow}(\lambda_1)$ are the probabilities in the reverse process to be in F at $\lambda_0$ conditioned to start in F(U) at $\lambda_1$. By definition $p^{F(U)}_{\leftarrow}(\lambda_1)=1$, because the system always ends in F at $\lambda_0$. All quantities in Eqs.(\ref{pFF},\ref{pFU}) are measured at the pulling rate $r_F$.

Analogously, the detailed work-FT in the range $\lambda_1 \rightarrow \lambda_2$,  is given by \cite{Juni2009}: 
\begin{subequations}
    \begin{align}
       \frac{p^F_{\rightarrow}(\lambda_1)} {p^F_\leftarrow(\lambda_1|r_F)} \frac{\rho_{{\lambda_1},{\lambda_2}}(W|F)}{\rho_{{\lambda_1},{\lambda_{2}}}(-W|F,r_F)} = \exp\left(\beta (W - \Delta G_{F,U}^{\lambda_1,\lambda_2})\right) \label{pUF} \\ \nonumber\\
           \frac{p^U_{\rightarrow}(\lambda_1)} {p^U_\leftarrow(\lambda_1,r_U)} \frac{\rho_{{\lambda_1},{\lambda_2}}(W|U)}{\rho_{{\lambda_1},{\lambda_{2}}}(-W|U,r_U)} = \exp\left(\beta (W - \Delta G_{U,U}^{\lambda_1,\lambda_2})\right)
           \label{pUU}
    \end{align}
\end{subequations} 
where $\Delta G_{F(U),U}^{\lambda_1,\lambda_2}$ is the free energy difference between state U at $\lambda_2$ and state F(U) at $\lambda_1$, $\rho_{\lambda_1,\lambda_2}(W|F(U))$ is the work distribution for the class of forward trajectories that start in F(U) at $\lambda_1$ and end in U at $\lambda_2$. $\rho_{\lambda_1,\lambda_2}(-W|F(U),r_{F(U)})$ is the corresponding reverse work distribution at the corresponding unloading rate, $r_F$ (Eq.\eqref{pUF}) and $r_U$ (Eq.\eqref{pUU}). $p^{F(U)}_{\rightarrow}(\lambda_1)$ are the probabilities in the forward process to be in U at $\lambda_2$ conditioned to start at F(U) in $\lambda_1$. By definition $p^{F(U)}_{\rightarrow}(\lambda_1)=1$, because the system always ends in U at $\lambda_2$. $p^{F(U)}_\leftarrow(\lambda_1|v_{F(U)})$ are the probabilities in the reverse process to be in F(U) at $\lambda_1$ conditioned to start in U at $\lambda_2$ with unloading rate $r_F$ ($r_U$). The unloading rate value for quantities in the reverse process are explicitly indicated in Eqs.(\ref{pFF},\ref{pFU}).  \\

From Eqs.(\ref{pFF}, \ref{pFU}, \ref{pUF}, \ref{pUU}) we calculate the partial forward work distributions across the whole range $\lambda_{min}\to\lambda_{max}$ for the two classes of trajectories:
\begin{subequations}
\begin{eqnarray}
 & & \rho_\rightarrow(W|F) = \int \rho_{\lambda_0,\lambda_1}(W_1|F) \rho_{\lambda_1,\lambda_2}(W_2|F) \cdot \nonumber\\
 & & \cdot \delta(W-W_1-W_2)dW_1dW_2 
\end{eqnarray}
\begin{eqnarray}
  & &\rho_\rightarrow(W|U) = \int \rho_{\lambda_0,\lambda_1}(W_1|U) \rho_{\lambda_1,\lambda_2}(W_2|U) \cdot  \nonumber \\
           & & \cdot \delta(W-W_1-W_2)dW_1dW_2 ~ .
\end{eqnarray}
\end{subequations} 
Putting everything together we obtain the detailed work-FTs for DTF.
\begin{subequations}
 \begin{eqnarray}
       & & \boxed{\frac{\rho_\rightarrow(W|F)}{\rho_\leftarrow(-W|F,r_F)} = \exp\left(\beta(W-\Delta G_{FU}+k_BTI_F)\right)} \nonumber \\
       & & \text{with } ~ \boxed{ I_F = \log \left( \frac{p_\leftarrow^F(r_F)} {p_\rightarrow^F} \right)} \label{detailedrhoF}
       \end{eqnarray}
\begin{eqnarray}
       & & \boxed{\frac{\rho_\rightarrow(W|U)}{\rho_\leftarrow(-W|U,r_U)} = \exp\left(\beta(W-\Delta G_{FU}+k_BTI_U)\right)} \nonumber \\
       & & \text{with } ~  \boxed{ I_U = \log \left( \frac{p_\leftarrow^U(r_U)} {p_\rightarrow^U} \right)}
           \label{detailedrhoU}
\end{eqnarray}
\end{subequations} 
where $\Delta G_{FU}=\Delta G_{F,U(F)}^{\lambda_0,\lambda_1}+\Delta G_{U(F),U}^{\lambda_1,\lambda_2}=G_U(\lambda_{max})-G_F(\lambda_{min})$ is the full free-energy difference and where we the argument $\lambda_1$ has dropped from the fractions $p^{F(U)}_{\rightarrow(\leftarrow)}(\lambda_1)$. Equations (\ref{detailedrhoF}, \ref{detailedrhoU}) are the detailed feedback-FTs reported in Eq.\eqref{eq.dFT-DTF} of the main text.

From Eqs.(\ref{detailedrhoF}, \ref{detailedrhoU}) we compute the full work-FT as follows:
\begin{eqnarray*}
\rho_\rightarrow(W)&=& \rho_\rightarrow(W|F)p_\rightarrow^F+\rho_\rightarrow(W|U)p_\rightarrow^U\\
&=&\rho_\leftarrow(-W)e^{\left( \beta\left(W - \Delta G_{FU} +k_BT\log\left[p^F_\leftarrow(r_F)+p^U_\leftarrow(r_U) \right]\right)\right)}
\label{fullrho}
\end{eqnarray*}
where:
\begin{eqnarray}\label{rhowreverse}
&&\rho_\leftarrow(-W) =\\
&&\frac{ \left( p_\leftarrow ^F(r_F) \rho _\leftarrow(-W|F,r_F)  + p_\leftarrow ^U (r_U) \rho _\leftarrow(-W|U,r_U) \right)}{(p_\leftarrow^F(r_F)+p_\leftarrow^U(r_U))}.\nonumber
\end{eqnarray}
Therefore we obtain the full work-FT for DTF, Eqs.(\ref{eq.FT-DTF}a, \ref{eq.FT-DTF}b) in the main text: 
\begin{eqnarray}
& &\boxed{\frac{\rho_\rightarrow(W)}{\rho_\leftarrow(-W)} = \exp\left(\beta (W-\Delta G_{FU} + k_BT\Upsilon_2)\right)} \nonumber \\
& & ~ {\rm with} ~ \boxed{\Upsilon_2 = \log\left(p^F_\leftarrow(r_F)+p^U_\leftarrow(r_U) \right)}
\end{eqnarray}
%

%/////////////////////////////////////////////////////////////////////

\section{The Bell-Evans model in single-hopping approximation}
\label{sec:BellEvans}
% ///////////////// Figure C1 : State vs force
\begin{figure}
    \centering
    \includegraphics{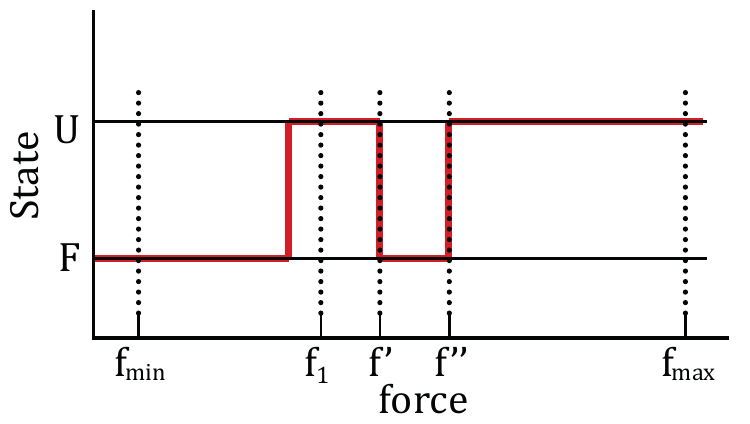}
    \caption{{\bf Single-hopping approximation.} In the model the molecule is initially folded (F) and pulled at $\dot{f} = r_F$ and the pulling rate is changed to $\dot{f}=r_U$ at a given force value ($f_1$) if the molecule is observed to be unfolded (U). In the single-hopping approximation the molecule refolds and unfolds again before $f_{max}$. The forces that determine the trajectories $U \rightarrow F \rightarrow U$ between $f_1$  and $f_{max}$ are $f'$ and $f''$, which are the folding and unfolding forces, respectively.}
    \label{fig9}
\end{figure}
% ____________________________________________________________________
%
% _____________________________________ FIGURE C2 
\begin{figure*}
\centering\includegraphics{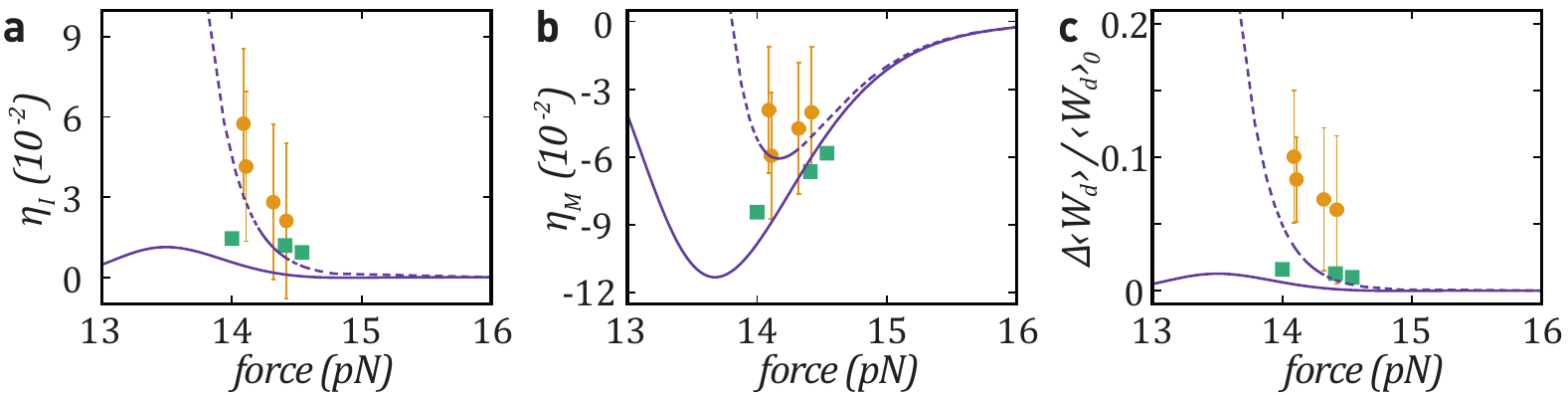} 
    \caption{{\bf The Bell-Evans model in the single-hopping approximation.} (a) Dissipation-reduction efficiency $\eta_I$ (b) Information-to-measurement efficiency $\eta_M$ and (c) Dissipation reduction versus force from the experiments (yellow circles), simulations (green squares), the exact single-hopping approximation solution Eq.\eqref{AWd_BE} (purple line) and the $1/r$ approximation Eq.\eqref{AWd_Hf} (dashed purple line). }\label{fig10}
\end{figure*}
% ____________________________________________________________________
To better understand under which conditions $\eta_I$ and $\eta_M$ are optimal we have carried out an analysis of DTF in the two-states Bell-Evans model where force is the control parameter. To analyze dissipation reduction we define $\Delta \langle W_d \rangle = \langle W_d \rangle_0 - \langle W_d \rangle > 0$, the change in the average dissipated work upon implementing feedback. In the Bell-Evans approximation where the force is controlled, has been shown to be qualitatively identical and quantitatively comparable to the experimental condition where the trap position is controlled \cite{Mossa2009}. In Figure \ref{fig9} we show a typical trajectory (state versus force) in the DTF protocol in the model where the initially folded (F) molecule is pulled at $\dot{f} = r_F$ and the pulling rate changed to $\dot{f} = r_U$ at a given force value ($f_1$) if the molecule is observed to be unfolded (U). $\Delta \langle W_d \rangle$ is only determined by the contribution of those trajectories that are in U at $f_1$: trajectories that are at F at $f_1$ do not change the pulling rate and therefore do not contribute to $\Delta \langle W_d \rangle$.
To determine $\Delta \langle W_d \rangle$ we restrict the analysis to single-hopping trajectories of the type $U\rightarrow F \rightarrow U$ after $f_1$. The average dissipated work in the range ($f_1,f_{max}$)  for the U-type trajectories is given by:
\begin{equation}\label{Wd_BE}
    \langle W_d \rangle_{U \rightarrow  F \rightarrow U} = P_{U \rightarrow  F \rightarrow U} (f_1 )\langle f''- f' \rangle x_m
\end{equation}
where $x_m$ stands for the difference in molecular extension between U and F; $f'$ and $f''$ are the folding and unfolding forces of steps $U\rightarrow F$ and $F\rightarrow U$ for the trajectory $U\rightarrow F \rightarrow U$ (Fig.\ref{fig9}); $P_{U\rightarrow F \rightarrow U} (f_1)$ is the fraction of trajectories of the type $U\rightarrow F \rightarrow U$, which in the current single-hopping approximation equals $1-P_s^U (f_1,f_{max})$ where $P_s^U (f_1,f)$ is the survival probability of U between $f_1$ and $f$. $\langle f''- f'\rangle x_m$ is given by,
\begin{eqnarray} \label{Af_BE}
& &\langle f''- f'\rangle =\\
& &-\int_{f_1}^{f_{max}} df' \frac{\partial P_s^U (f_1,f')}{\partial f'} \int_{f'}^{f_{max}} df'' P_s^F (f',f'')\nonumber
\end{eqnarray}
where $P_s^F (f',f'')$ is the survival probability of F between $f'$ and $f''$. 
$\Delta \langle W_d \rangle$ is proportional to the difference of the average dissipated work between $f_1$ and $f_{max}$ calculated at the pulling rates $r_U$ and $r_F$. Equations \ref{Wd_BE},\ref{Af_BE} must be calculated at the pulling rates $r_F$ and $r_U$ to obtain the dissipation reduction in the \textit{single-hopping} approximation,
\begin{eqnarray}\label{AWd_BE}
&&\Delta \langle W_d \rangle =\\ &&p_{\rightarrow}^U(r_F) \left[ \langle W_d \rangle_{U \rightarrow  F \rightarrow U}(r_U) - \langle W_d \rangle_{U \rightarrow  F \rightarrow U}(r_F) \right]\nonumber
\end{eqnarray}
where $p_\rightarrow^U (r_F)$ is the fraction of trajectories observed at U in $f_1$ starting at F in $f_{min}$ and the dissipated work is restricted to the range ($f_1,f_{max}$). For practical purposes we can take $f_{max}\rightarrow \infty$ as the molecule always ends in U at $f_{max}$. Equations \ref{Wd_BE},\ref{Af_BE},\ref{AWd_BE} can be numerically calculated for generic Bell-Evans rates where survival probabilities have simple analytical expressions. For the specific case relevant to the experiments (L4 molecule) of a transition state located at half distance between F and U, $x^\dagger = x_m/2$, we have
\begin{subequations}\label{kinetics_BE}
\begin{align}
k_{F\rightarrow U}(f) = k_0 \cdot \exp\left( \beta fx^\dagger\right) \\
k_{F\leftarrow U}(f) = k_0 \cdot \exp\left[ \beta( \Delta G - fx^\dagger)\right]
\end{align}
\end{subequations}
where $k_0$ is the kinetic rate at zero force and $\Delta G$ is the free energy difference between F and U ($\beta = 1/k_B T$). Finally, for sufficiently high forces where $k_{F\leftarrow U} (f_1)/(\beta x^\dagger r)\ll 1$, the average dissipation reduction is obtained to first order in $1/r$:
\begin{equation}\label{AWd_Hf}
\Delta \langle W_d \rangle =  p_{\rightarrow}^U(r_F) \left[ \frac{x_m k_{F\leftarrow U}(f_1)^2}{2k_{F\rightarrow U}(f_1) (\beta x^\dagger)^2} \right](\frac{1}{r_F} - \frac{1}{r_U}) 
\end{equation}
which is positive for $r_U>r_F$ as expected, and negative otherwise. For practical purposes we take $f_{min}\rightarrow -\infty$ as the molecule always starts in F at $f_{min}$. $p_{\rightarrow}^U (r_F)$ is expressed as:
\begin{equation}\label{pU_rF_U}
    p_\rightarrow^U (r_F)= 1 - \exp  \left( -\frac{k_{F\rightarrow U}(f_1)}{\beta x^\dagger r_F } \right)
\end{equation}
To calculate the efficiencies, we also need the values of $\langle W_d \rangle_0$ and $k_B T\Upsilon_2$ for DTF, the latter being given by Eq.(\ref{eq.FT-DTF}b).  $\langle W_d \rangle_0$ is estimated from the mean first unfolding force $\langle f_{F\rightarrow U}\rangle$ in the Bell-Evans approximation,
\begin{eqnarray}\label{AWd0_BE}
\langle W_d \rangle_0 = x_m \left( \langle f_{F \rightarrow U} \rangle - f_c\right) = \nonumber \\
= x_m \left(\frac{1}{\beta x^\dagger} \log \left( \frac{\beta x^\dagger r}{k_0}\right) - f_c\right)
\end{eqnarray}
where $f_c=\Delta G / x_m$  is the coexistence force and $k_{F\rightarrow U}(f_c)=k_{F\leftarrow U}(f_c)=k_c$ is the rate at coexistence. We also have,
\begin{subequations} \label{kTY2_pU}
\begin{align}
    k_B T\Upsilon_2 = \log ( p_\leftarrow^F (r_F) + p_\leftarrow^U (r_U)) \\
    p_{\leftarrow}^U(r) = \exp\left( -\frac{k_{F \leftarrow U}(f)}{\beta x^\dagger r} \right)
\end{align}
\end{subequations}
and $p_\leftarrow^F(r)=1-p_\leftarrow^U (r)$. We have calculated these quantities for the parameters that fit the experimental pulling curves for L4 without feedback  ($k_0=2\cdot 10^{-14} s^{-1}$, $x_m = 18 nm$, $x^\dagger=9 nm$, $k_B T=4.114 pN\cdot nm$, $\beta = 1/k_B T\sim 0.24 pN^{-1}\cdot nm^{-1}$, $\Delta G=264.0 pN\cdot nm$, $f_c = \Delta G ⁄ x_m = 14.66 pN$, $k_c=1.7 s^{-1}$. From these values we calculate the efficiencies defined in Eqs.(\ref{eq.etaI},\ref{eq.etaM}),
\begin{subequations} \label{effi_BE}
\begin{align}
    \eta_I = \frac{\Delta \langle W_d \rangle }{\langle W_d \rangle_0 + k_BT \Upsilon_2} \\
    \eta_M = \frac{\Delta \langle W_d \rangle - k_BT \Upsilon_2}{\langle W_d \rangle_0}
\end{align}
\end{subequations}
%
% _____________________________________ FIGURE C3 
\begin{figure*}
\centering\includegraphics{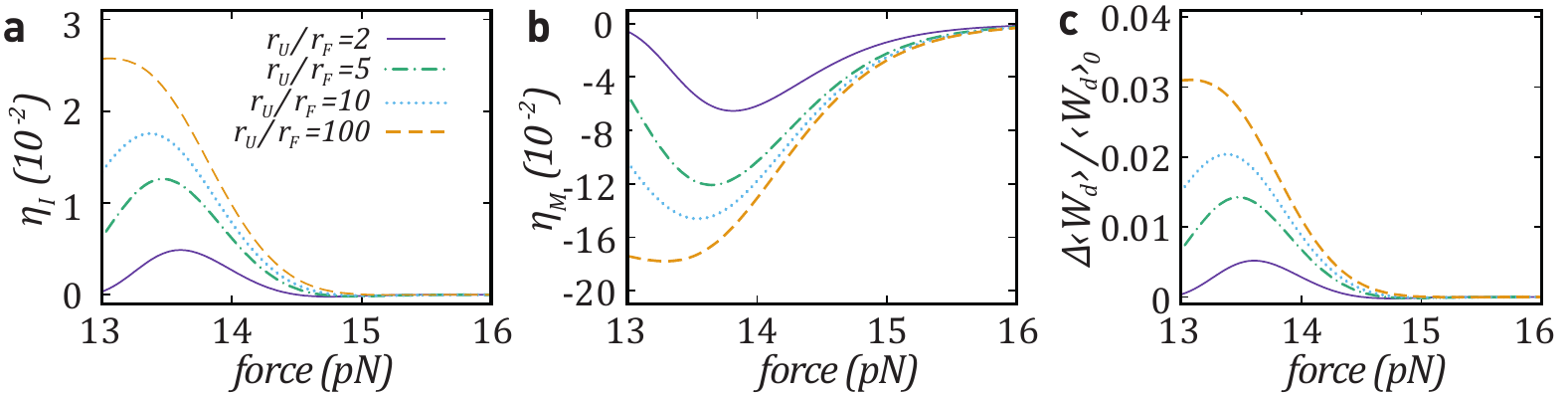}
    \caption{{{\bf Bell-Evans model prediction upon varying $r_U/r_F$} (a) Dissipation-reduction efficiency $\eta_I$ (b) Information-to-measurement efficiency $\eta_M$ (c) Dissipation reduction versus force using the single-hopping approximation solution.}}
    \label{fig11}
\end{figure*}
% ____________________________________________________________________
In Figure \ref{fig10} we show $\eta_I$ and $\eta_M$ versus force for $r_F=4$pN/s and $r_U=17$pN/s. The purple continuous line shows the results obtained from Eqs.(\ref{effi_BE}a,b) by numerical calculation of $\Delta \langle W_d \rangle$ in the single-hopping approximation Eqs.(\ref{Wd_BE},\ref{Af_BE},\ref{AWd_BE}). The dashed line shows the $1/r$ leading term, Eq.\eqref{AWd_Hf}, which holds for sufficiently high forces ($f>14.45$pN where $k_{F\leftarrow U}(f)/(\beta x^\dagger r_F)\leq 0.3$). As we can see from the figure while $\eta_I$ is small and positive ($\sim10^{-2}$), $\eta_M$ is small and negative ($\sim-10^{-2}$) in the experimentally measured range of forces. The points are the experimental results shown in Figure \ref{fig3}b.

In Figure \ref{fig10} we show the analytical Bell-Evans predictions for $\eta_I$ and $\eta_M$ in the single-hopping approximation for a fixed $r_F$ and varying $r_U$. Interestingly, while $\Delta \langle W_d \rangle$ and $\eta_I$ increase with $r_U$, the behavior of $\eta_M$ is the opposite, becoming more negative as $r_U$ increases. 

% ___________________________________________________________
\bibliography{bibliography}% Produces the bibliography via BibTeX.

\end{document}

\maketitle

\tableofcontents
 
%\appendix
\setcounter{section}{0}
\setcounter{figure}{0}
\setcounter{equation}{0}
\makeatletter 
\renewcommand{\thefigure}{S\@arabic\c@figure}
\makeatother
\makeatletter 
\renewcommand{\theequation}{S\@arabic\c@equation}
\makeatother
\makeatletter
\renewcommand{\thetable}{S\@arabic\c@table}
\makeatother

\newpage

%//////////////////////////////////////////////////////////////////////////////
\newpage
\section{Experimental measurement of $\Delta G_{FU}$ without feedback}\label{SI:AGFU}
% _____________________________________ FIGURE 1 
\begin{figure}[H]
    \centering
    \includegraphics[width = 0.8\textwidth]{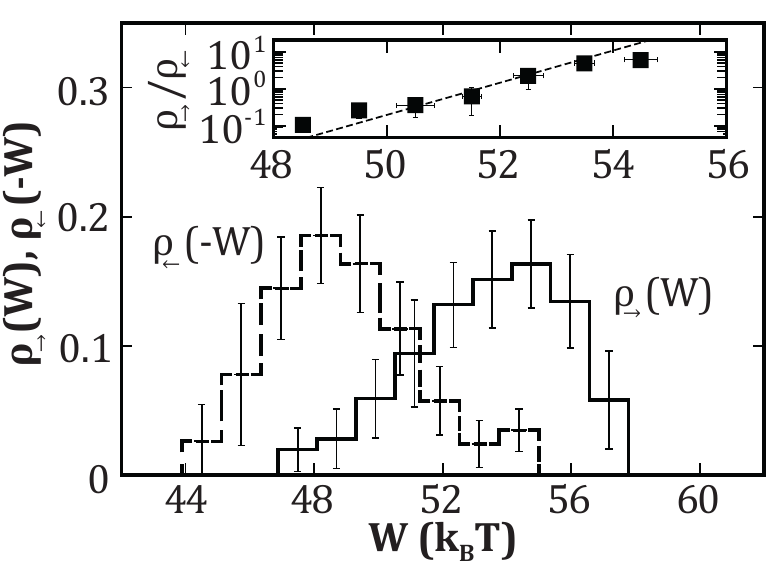}
    \caption{{\bf Work-FT results without feedback.} In the main plot the forward (solid black line) and reverse (dashed black line) work distributions without feedback for the DNA hairpin pulled at the loading rate $r=4$pN/s. In the inset we show the test of the Crooks-FT ($\rho(W)/\rho(-W) = \exp ((W-\Delta G_{FU})/k_BT)$) from which we extract the value of $\Delta G_{FU}$. We obtain $\Delta G_{FU} = 51\pm1  ~{\rm k_BT}$.}
    \label{figS1}
\end{figure}
% ___________________________________________________________________

%//////////////////////////////////////////////////////////////////////////////
\newpage
\section{Simulation details}\label{SI:details}
To simulate pulling experiments where the trap position, $\lambda$, is the control parameter we proceed as described in references \cite{Evans2001,Mano2005}. Briefly, the forward and reverse trajectories are simulated as a first order Markov chain where, for a given value of $\lambda$ and force, $f$, it is satisfied that:
\begin{equation}
    \lambda(f) = x_{DNA}^\sigma(f) + x_h(f) + x_b(f)~.
\label{distance}
\end{equation}
In Eq.\eqref{distance} $x_b(f)$ is the position of the trapped bead relative to the center of the optical trap; $x_h(f)$ is the end-to-end distance of the dsDNA handles and $x_{DNA}^\sigma$ is the end-to-end distance of the DNA hairpin, which depends on the state of the molecule ($\sigma = F,U$). The latter defines two force branches, one when the DNA hairpin is folded and the other when the hairpin is unfolded. 

To determine the two force branches, first we need to determine the extension of each element at a given $\lambda$, i.e., at a given force. On one hand, $x_b(f)$ satisfies:
\begin{equation}
    |f| = k_b x_b
\end{equation}
being $k_b$ the trap stiffness ($k_b = 0.068~{\rm pN/nm}$). In this simple approximation (Hooke's law), the bead behaves as a Brownian particle under the action of two opposing springs (the optical trap and the molecular construct), hence the white noise due to thermal fluctuations satisfies:
\begin{equation}
\langle \delta x^2 \rangle = \frac{k_BT}{k_{mol}^\sigma+k_b} ~ , ~ ~ ~ ~  \langle \delta f^2 \rangle = \frac{k_BT k_b^2}{k_{mol}^\sigma + k_b}   \label{variances}
\end{equation}
where $k_{mol}^\sigma$ is the stiffness of the molecular construct. It is given by $1/k_{mol}^\sigma = 1/k_h + 1/k_{DNA}^\sigma$) where $k_h$ and $k_{DNA}$ are the stiffness of the dsDNA handle and the DNA hairpin. These stiffness values are calculated as the derivative of the force with respect to the extension of the corresponding element (handle or molecule). Furthermore, the elastic response of the dsDNA handles and the unfolded DNA hairpin is well described by the Worm-Like Chain model,
\begin{equation}
    f_{\rm WLC} = \frac{k_BT}{4L_p} \left (\left( 1-\frac{x}{L_c}\right)^{-2} +4\frac{x}{L_c} -1 \right)
\end{equation}
where $k_B$ is the Boltzmann's constant and $T$ is the temperature. $f_{\rm WLC}$ is the measured force at a given end-to-end extension of the dsDNA handles (29 base-pairs at each flanking side) and the unfolded hairpin, $x$; $L_p$ is the persistence length ($L_p = 10\rm nm$ for the dsDNA handles and $L_p = 1.34\rm nm$ for the unfolded DNA hairpin \cite{Evans2001,Mano2005,Mano2007}) and $L_c$ is the contour length ($L_c = 58{\rm bp}\cdot0.34{\rm nm/bp}=19.72\rm nm$ for the dsDNA handles and $L_c = 25.96\rm nm$ for the unfolded DNA hairpin \cite{Evans2001,Mano2005,Mano2007}). Besides, the folded DNA hairpin behaves as a dipole of length ($d_0 = 2 \rm nm$ \cite{Mano2007}) of extension:
\begin{equation}
    x_d(f) = d_0\left[ \coth\left( \frac{d_0 f}{k_BT} \right) - \frac{k_BT}{d_0 f}\right]\label{dipole}
\end{equation}
 To simulate pulling experiment, we calculate the unfolding, $k_\rightarrow(\lambda)$, and refolding, $k_\leftarrow(\lambda)$, kinetic rates as described in \cite{Ale2014}. The kinetic rates are defined as:
\begin{subequations}
    \begin{align}
k_{\rightarrow}(\lambda) &= k_0 \exp\left( - \frac{B(\lambda)}{k_BT}\right) \\
k_{\leftarrow}(\lambda) &= k_0 \exp\left( - \frac{B(\lambda) - \Delta G_{FU}(\lambda)}{k_BT}\right)
    \end{align}
\end{subequations}
where $B(\lambda)$ is the kinetic barrier, $k_0$ is the attemp rate ($k_0 = 8 \cdot 10^4 s^{-1}$) and $\Delta G_{FU}(\lambda)$ is the energy difference between states F and U at a given $\lambda$. The latter is determined as:
\begin{equation}
    \Delta G_{FU}(\lambda) = \Delta G_{FU} + \Delta W_{FU}^{DNA} + \Delta W_{FU}^h + \Delta W_{FU}^b
\end{equation}
where $\Delta G_{FU}$ is the folding free energy of the DNA hairpin ($\Delta G_{FU} = 51 ~{\rm k_BT}$); $\Delta W_{FU}^{DNA}$ is the stretching contribution of the unfolded hairpin ($W_U^{DNA} = \int_0^{x_{DNA}^U} f_{\rm WLC}(x')dx'$) minus the orientation contribution of the folded DNA hairpin ($W_F^{DNA} = \int_0^{x_{DNA}^F} f_{\rm d}(x')dx'$) with $f_d(x)$ the inverse function of Eq.\eqref{dipole}; $\Delta W_{FU}^h$ is the stretching contribution of the handles ($\Delta W_{FU}^{h} = \int_{x_h^F}^{x_h^U} f_{\rm WLC}(x')dx'$) where the limits are the end-to-end distances at a given $\lambda$ evaluated at the different forces, when the hairpin is folded and unfolded; $\Delta W_{FU}^b$ is the work done to displace the bead with respect to the center of the optical trap between the folded and unfolded branches ($\Delta W_{FU}^{b} = \int_{x_b^F}^{x_b^U} f_{\rm Hooke}(x')dx'$).

Finally the kinetic barrier $B(\lambda)$ can be simplified using the Bell-Evans approximation. However, as we want to see if this model fits our experiments we calculated explicitly the kinetic barrier using the Kramers solution to the one-dimensional diffusion problem in the unzipping molecular free energy landscape described by the nearest-neighbour model \cite{Ale2014,Kos2015}:
\begin{equation}
    B(\lambda) = k_BT \log\left( \sum_{n=0}^N \sum_{n'=0}^n \exp\left[ \frac{\Delta G_n(\lambda) - \Delta G_{n'}(\lambda)}{k_BT} \right]\right)~ .
    \label{Beff}
\end{equation}
In Eq.\eqref{Beff} $\Delta G_n(\lambda)$ is the free energy of the hairpin determined by the number of opened base-pairs, $n$, at a given $\lambda$. This energy is calculated as
\begin{equation}
    \Delta G_n(\lambda) = \Delta G_n + \int_0^{x_n^{DNA}}f_{\rm WLC}(x')dx' +  \int_0^{x_h^n}f_{\rm WLC}(x')dx' + \int_0^{x_b^n}f_{\rm Hooke}(x')dx' ~ .
\end{equation}
Here, $x_n^{DNA}$ is the extension of DNA when only $n$ base-pairs are opened, $x^n_h$ and $x_b^n$ are the extension of the handles and the bead position when only $n$ base-pairs are opened.

The simulated trajectories are generated as follows: the forward trajectory is initialized at the folded state ($n = 0$) at $\lambda_{min} = 0{\rm nm}$, while the reverse one is initialized at the unfolded state ($n = N$) at $\lambda_{max} = 300{\rm nm}$. During the simulation the relaxation of the handles and the bead is assumed to be instantaneous. The steps of the algorithm for the forward (reverse) process are:
\begin{enumerate}
% Primer pas ____________________________
\item { $\lambda$ increases (decreases) by the amount $r\Delta t/k_b$, were $r$ is the loading rate and $\Delta t$ is the inverse of the data acquisition frequency ($\Delta t = 0.001$s).}
% Segon pas _____________________________
\item {Eq.\eqref{distance} is solved according to the state of the hairpin to find the value of the force $f$ acting on the experimental setup. Moreover, to be more realistic we added a Gaussian noise of zero mean and variance given by Eq.\eqref{variances} to the measured $\lambda$ and $f$.}
% Tercer pas _____________________________
\item {We calculate the probability to observe an unfolding or folding transition as: $k_{\rightarrow}(\lambda)\cdot\Delta t$ if the molecule is folded or as $k_{\leftarrow}(\lambda)\cdot\Delta t$ if the molecule is unfolded. }
% Quart pas _____________________________
\item {We compare the transition probability with an uniformly distributed random number between 0 and 1. If the probability is larger than the random number we change the state of the molecule in the next step, otherwise the molecule remains in the same state.  }
\end{enumerate}
The work done in the simulated forward and reverse trajectories is calculated as the area under the trajectory, i.e. $W = \int_{\lambda_{min}}^{\lambda_{max}}f(\lambda') d\lambda'$.

%//////////////////////////////////////////////////////////////////////////////
\newpage
\section{Numerical simulation of hairpins L4 and L8}\label{SI:simuL4L8}

% _____________________________________ FIGURE 2 
\begin{figure}[H]
    \centering
    \includegraphics{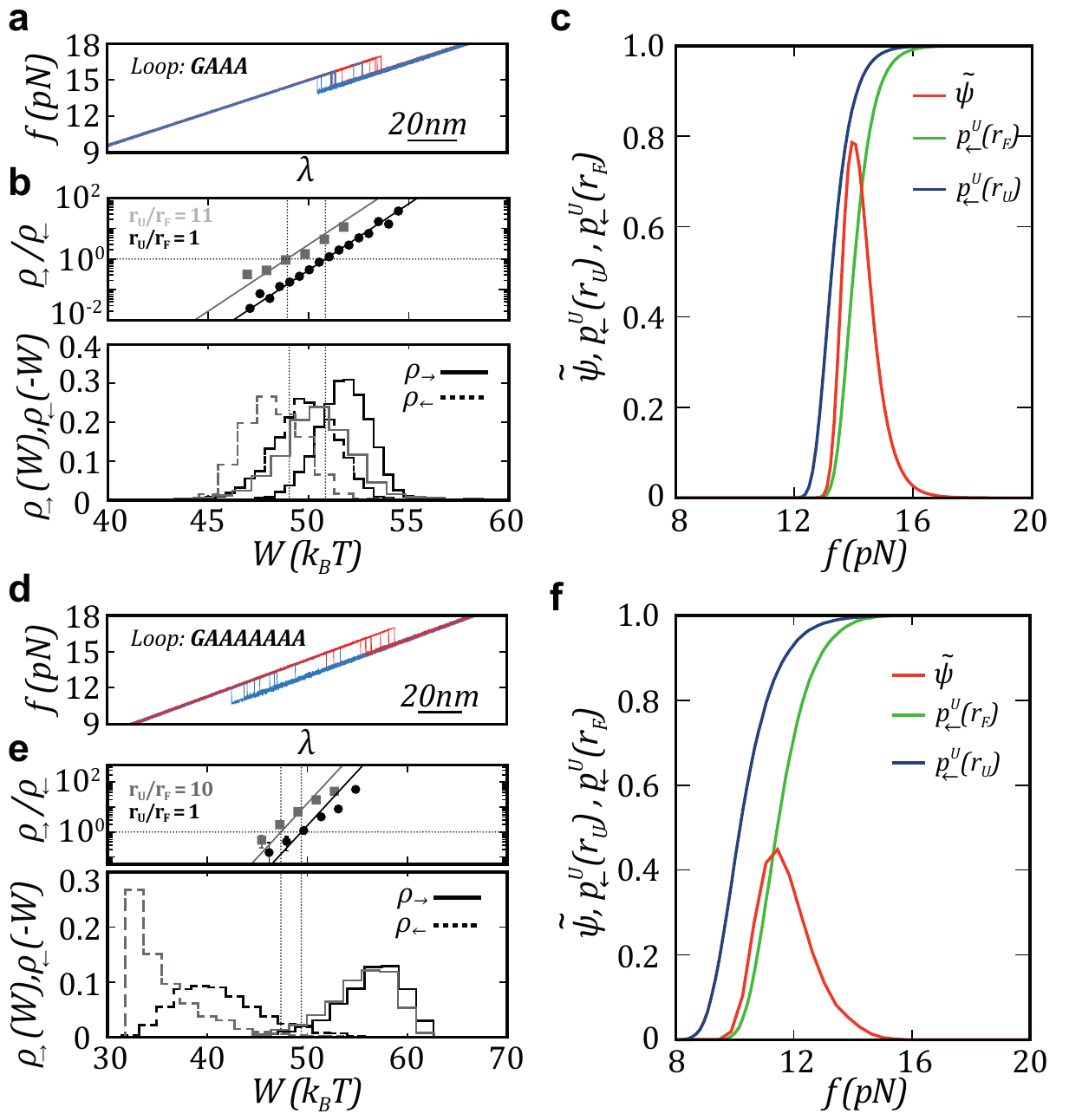}
    \caption{{\bf Numerical simulation of hairpins L4 and L8.} (a) Simulated forward (red) and reverse (blue) FDCs without feedback for the DNA L4 hairpin. (b) Bottom: Forward work distributions (solid lines) and properly weighed reverse work distributions (dashed lines) for the non-feedback case (black lines) and the CTF with $r_U/r_F = 11$ (gray lines). Top: Test of the feedback-FT (Eq.16c) for the cases $r_U/r_F = 1$ (black circles) and $r_U/r_F = 11$ (gray squares) (c) $p_\leftarrow^U\left(\lambda,r_U\right)$, $p_\leftarrow^U\left(\lambda,r_F\right)$ and last-folding density $\widetilde{\psi}(\lambda)$ for $r_F = 5$pN/s and $r_U = 20$pN/s at different measuring points. (d)  Simulated forward (red) and reverse (blue) FDCs without feedback for the DNA L8 hairpin. (e) Bottom: Forward work distributions (solid lines) and properly weighed reverse work distributions (dashed lines) for the non-feedback case (black lines) and the CTF with $r_U/r_F = 10$ (gray lines). Top: Test of the feedback-FT (Eq.16c) for the cases $r_U/r_F = 1$ (black circles) and $r_U/r_F = 10$ (gray squares). (f) $p_\leftarrow^U\left(\lambda,r_U\right)$, $p_\leftarrow^U\left(\lambda,r_F\right)$ and last-folding density $\widetilde{\psi}(\lambda)$ for $r_F = 6$pN/s and $r_U = 30$pN/s at different measuring points. In panels (c) and (f) we used force as a reference value to present the results. Force is more informative than the trap position $\lambda_1$ because the latter is defined as the relative distance between the trap position and an arbitrary initial position in the light-lever detector. }
    \label{figS2}
\end{figure}
% ___________________________________________________________________

%//////////////////////////////////////////////////////////////////////////////
\newpage
\section{Bias of the DTF+CTF strategy}\label{SI:DTFCTF}

% _____________________________________ FIGURE 3
\begin{figure}[H]
    \centering
    \includegraphics[width = \textwidth]{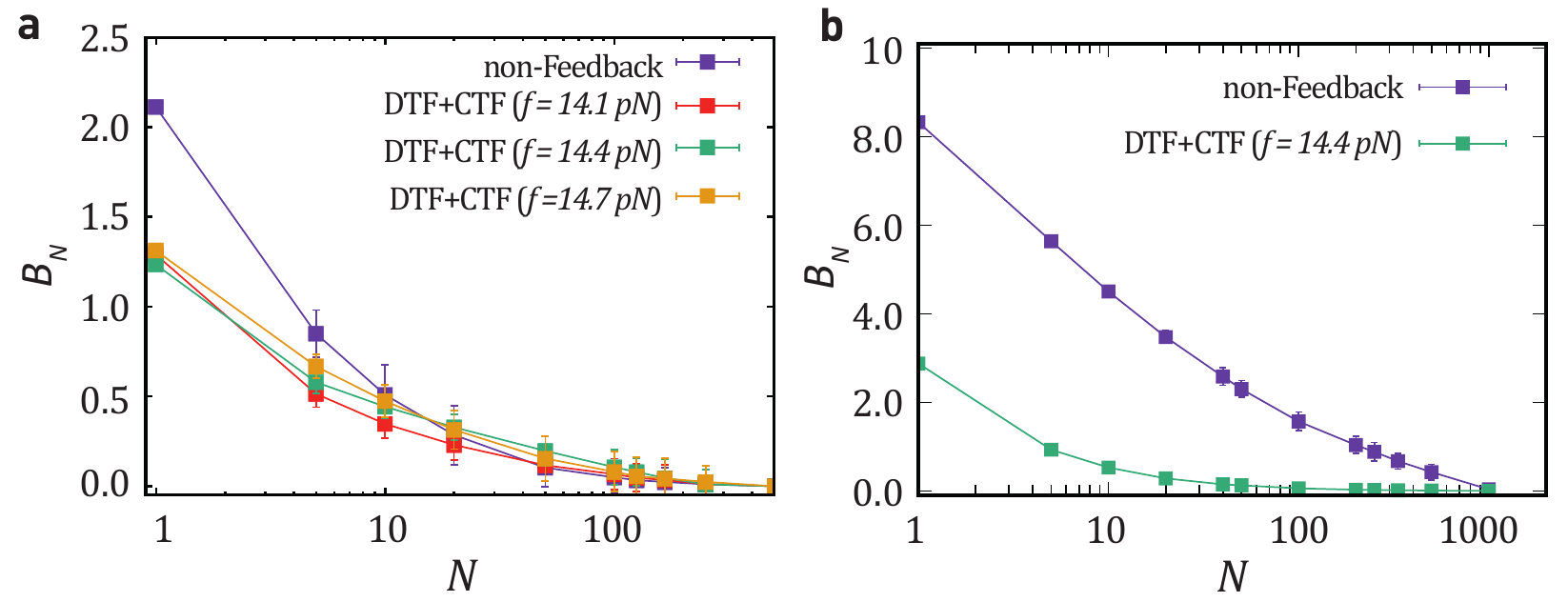}
    \caption{{\bf Bias for the numerical simulation of the DTF+CTF strategy.} a)  Bias for the DTF+CTF strategy using molecule L4. The molecule is initially pulled at $r_F=4$pN/s with DTF until $\lambda_1$, where an observation is made. If the outcome is U then the pulling rate is switched to $r_U = 17$pN/s between $\lambda_1$ and $\lambda_{max}$. Instead, if the outcome is F the pulling rate is reduced to $r_F'=1$pN/s $< r_F$ and the CTF protocol turned on. In this case, at the first unfolding event after $\lambda_1$, the pulling rate is switched to $r_U> r_F>r_F'$ until $\lambda_{max}$. In the DTF+CTF strategy both U- and F-trajectories contribute to dissipated work reduction. In the three different $\lambda_1$ the $\langle W_d\rangle$ is reduced by $\sim 1k_BT$ respect the non-feedback case. b) Bias for DTF+CTF strategy applied to molecule L8 using the same pulling speeds than the ones used in panel a). In this case, for the studied specific conditions, the $\langle W_d\rangle$ is reduced by $\sim 6k_BT$ respect the non-feedback case. }
    \label{figS3}
\end{figure}
% ___________________________________________________________________

\printbibliography